\documentclass[iop]{emulateapj}
\usepackage{apjfonts}
\pdfoutput=1

\newcommand{\ha}{\rm{H}\alpha}
\newcommand{\hb}{\rm{H}\beta}
\newcommand{\mic}{\mu{\rm m}}
\newcommand{\oh}{12+log(O/H)}
\newcommand{\galex}{{\it GALEX}}
\newcommand{\wise}{{\it WISE}}
\newcommand{\mzs}{{$M_*$--$Z$--SFR}}

\shorttitle{MASS-METALLICITY-SFR RELATION}
\shortauthors{SALIM ET AL.}

\begin{document}

\title{A Critical Look at the Mass--Metallicity--Star Formation Rate
  Relation in the Local Universe. I. An Improved Analysis Framework
  and Confounding Systematics}

\author{Samir Salim\altaffilmark{1}, Janice C.\ Lee\altaffilmark{2,3},
  Chun Ly\altaffilmark{4,5}, Jarle Brinchmann\altaffilmark{6}, Romeel
  Dav\'e\altaffilmark{7}, Mark Dickinson\altaffilmark{8}, John J.\
  Salzer\altaffilmark{1}, St\'ephane Charlot\altaffilmark{9}}
\altaffiltext{1}{Department of Astronomy, Indiana University,
  Bloomington, IN 47404, USA} 
\altaffiltext{2}{Space Telescope Science Institute, Baltimore, MD 21218, USA} 
\altaffiltext{3}{Visiting Astronomer, Spitzer Science Center, Caltech,
  Pasadena, CA 91125, USA}
\altaffiltext{4}{NASA Goddard Space Flight Center, Greenbelt, MD 20771, USA}
\altaffiltext{5}{NASA Postdoctoral Fellow}
\altaffiltext{6}{Leiden Observatory, Leiden University, NL-2300 RA Leiden, the Netherlands}
\altaffiltext{7}{University of the Western Cape, Bellville, Cape Town
  7535, South Africa}
\altaffiltext{8}{National Optical Astronomy Observatory, Tucson, AZ
  85719}
\altaffiltext{9}{Institut d'Astrophysique de Paris, CNRS, F-75014 Paris, France}
\email{salims@indiana.edu}

\begin{abstract}
  It has been proposed that the (stellar) mass--(gas) metallicity
  relation of galaxies exhibits a secondary dependence on star
  formation rate (SFR), and that the resulting \mzs\ relation may be
  redshift-invariant, i.e., ``fundamental.'' However, conflicting
  results on the character of the SFR dependence, and whether it
  exists, have been reported. To gain insight into the origins of the
  conflicting results, we (a) devise a non-parametric,
  astrophysically motivated analysis framework based on the offset
  from the star-forming (``main'') sequence at a given $M_*$ (relative
  specific SFR), (b) apply this methodology and perform a
  comprehensive re-analysis of the local \mzs\ relation, based on
  SDSS, {\it GALEX}, and {\it WISE} data, and (c) study the impact of
  sample selection, and of using different metallicity and SFR
  indicators. We show that metallicity is anti-correlated with
  specific SFR regardless of the indicators used.  We do not find that
  the relation is spurious due to correlations arising from biased
  metallicity measurements, or fiber aperture effects.  We emphasize
  that the dependence is weak/absent for massive galaxies ($\log
  M_*>10.5$), and that the overall scatter in the \mzs\ relation does
  not greatly decrease from the $M_*$--$Z$ relation.  We find that the
  dependence is stronger for the highest SSFR galaxies above the
  star-forming sequence. This two-mode behavior can be described with
  a broken linear fit in 12+log(O/H) vs. log (SFR$/M_*$), at a given
  $M_*$. Previous parameterizations used for comparative analysis with
  higher redshift samples that do not account for the more detailed
  behavior of the local \mzs\ relation may incorrectly lead to the
  conclusion that those samples follow a different relationship.
  
\end{abstract}


\section{Introduction}

The chemical enrichment of galaxies and its change through cosmic time
represents one of the key aspects of efforts to arrive at a
comprehensive picture of galaxy evolution. Therefore, it is of
particular importance that there may exist a relation that connects
the gas-phase metal abundance ($Z$, ``metallicity''), the stellar mass
of the galaxy ($M_*$), and its current star formation rate (SFR)
\citep{ellison08}, and that this relation may be redshift-independent,
or ``fundamental'' (hence, ``fundamental metallicity relation'', or
FMR, \citealt{mannucci10}), even though each of the quantities itself
evolves for a given galaxy. Despite an impressive amount of work
carried out in recent years, there remain fundamental uncertainties
concerning the empirical properties of the $M_*$--$Z$--SFR relation
\citep{yates12,andrews13}, its redshift invariance (e.g.,
\citealt{maier14,steidel14}) and even whether it exists
\citep{sanchez13}. This paper explores how the choice and treatment of
the observational data affect the perceived character of the
$M_*$--$Z$--SFR relation, and presents methodological recommendations
for a consistent approach to study it, and some findings based on the
application of this methodology.

The $M_*$--$Z$--SFR relation represents an extension of the
mass-metallicity ($M_*$--$Z$) relation (MZR). MZR was first studied
in a small sample of irregular galaxies \citep{lequeux79}, and was
later firmly established using much larger samples from the Sloan
Digital Sky Survey (SDSS) spectroscopic survey (\citealt{tremonti04},
hereafter T04). T04 found MZR to be more fundamental then the
previously studied luminosity-metallicity relations (e.g.,
\citealt{garnett02,lee04,salzer05}). The sense of the MZR is that more
massive galaxies have, on average, higher metallicities. The observed
scatter of MZR ($\sim$0.1 dex in metallicity) is usually described as
``tight'', although it should be kept in mind that the full range of
metallicities between gas-rich dwarfs ($\log M_*<8$)\footnote{Masses
  are expressed in units of solar mass ($M_{\odot}$).} and the most
massive star-forming (SF) galaxies ($\log M_*\approx 11$) is less than
a decade \citep{andrews13}.

Subsequently, building on efforts to study non-local
luminosity-metallicity relation (e.g., \citealt{kk04}), the MZR was
observed at intermediate redshifts (e.g.,
\citealt{savaglio05,cowie08,lamareille09,zahid11,cresci12,perez-montero13,stott13,ly14,mithi})
and is also starting to be measured at higher redshifts
($z\gtrsim1.6$), either from direct observations
\citep{maiolino08,zahid14a,tronosco14,steidel14,maier14}, stacked
spectra \citep{erb06,henry13,cullen14,yabe14}, or by exploiting
gravitational lensing
\citep{richard11,christensen12,wuyts12,belli13,yuan13}. In addition to
observational challenges, with some of the diagnostic lines redshifted
into the near-infrared region, non-local studies have to contend with
the changing characteristics of the interstellar medium (ISM)
\citep{nakajima14,steidel14} and the increased uncertainties regarding
the removal of galaxies in which AGNs contribute to ionization
\citep{kewley13,juneau14}, and sample selection effects
\citep{juneau14}. Nevertheless, the general consensus is that average
metallicities at a given mass were lower at higher redshifts. Note
that for a given galaxy the evolution in metallicity is even greater
than the offset between MZRs, because the galaxies we see today had a
smaller mass in the past.

The search for secondary dependencies of MZR can be traced back to
T04, who remarked that the scatter in MZR is approximately twice the
estimated error in metallicity, suggesting that other galaxy
properties may contribute to it. They found mass-dependent MZR
residuals with respect to the mass surface density, but not with
respect to the $\ha$ equivalent width, a rough proxy for specific star
formation rate (SSFR = SFR$/M_*$). Subsequently, also using SDSS, but
with different method of deriving metallicities, \citet{ellison08}
found that MZR residuals at a given mass do depend on the SSFR, and,
even more strongly, on galaxy's physical size. Both effects were found
to be more pronounced at lower masses. \citet{mannucci10} (hereafter
M10) introduced an analytical form for \mzs\ relation, and more
importantly, proposed that the \mzs\ relation, unlike the MZR, does
not evolve with redshift (hence, it is ``fundamental''). Thus, MZRs at
high redshift simply represent slices of the \mzs\ relation that can
be defined with the local data. M10 also found a projection of the
\mzs\ relation along the axis that lies in the $M_*$--SFR plane that
minimizes the scatter in metallicity compared to the MZR projection. A
similar concept, sharing the idea of tying together MZRs at different
redshifts, was concurrently put forward by \citet{lara-lopez10}
(hereafter LL10).

The idea of a fundamental relation was foreshadowed by \citet{hoopes},
who showed that UV luminous compact galaxies, which may represent
local lyman-break galaxy (LBG) analogs, lie on a MZR that is offset
from the general MZR towards lower metallicities. They noted that the
MZR of UV luminous compact galaxies resembles the $z\sim 0.7$ MZR of
\citet{savaglio05} in terms of the magnitude of the offset from the
local MZR and even the fact that the offset was greater for low-mass
galaxies. The invariant aspect of the \mzs\ relation was considered,
but dismissed by \citet{ellison08}, because they concluded that for
high-redshift galaxies the effect of higher SSFRs would be mostly
countered by the effect of high-redshift galaxies being smaller (size
was found by these authors to be a stronger driver of metallicity
dependence). However, nearly all subsequent studies have only
considered the effect of the evolving SFRs, but not also of the size.

The existence of \mzs\ relation and its possible fundamental aspect
were quickly adopted, and theoretical models, both analytical (e.g.,
\citealt{lilly13}) and numerical (e.g., \citealt{dave11}), were
developed to explain it (see Section 5). FMR has even been used as a
assumption to yield other predictions \citep{peeples13}. Furthermore,
there are numerous efforts to detect secondary dependencies even at
higher redshifts (e.g.,
\citealt{cresci12,zahid14a,wuyts14,steidel14,maier14,mithi}). Nevertheless,
the observational status of the \mzs\ relation, and its exact
character, are currently unclear even in the local universe where the
data are most abundant. For example, in their study of SDSS samples ,
\citet{yates12} find that the anti-correlation between SFR and
metallicity at a given massis only present at lower masses ($\log
M_*\lesssim10.2$), but then {\it reverses}, so that the metallicity is
{\it higher} for high-SFR galaxies, contradicting the results of
M10. More disturbingly, \citet{sanchez13} find no dependence of
metallicity on SFR at all in their sample of local galaxies with
spatially resolved metallicities. In contrast, \citet{andrews13}, by
measuring direct metallicities on stacked SDSS spectra, find not only
that metallicity is anti-correlated with SFR at all masses, but that
this dependence is two to three times stronger than that found by M10.

There are also open questions regarding what fraction of metallicity
scatter can be accounted for by allowing the SFR dependence, with
estimates ranging from very moderate \citep{ellison08,perez-montero13}
to quite substantial \citep{mannucci10}. Finally, the existence of
another ``second'' parameter on which metallicity may depend, the
galaxy size, has received relatively little attention, despite the
initial claims that it is even more important than the dependence on
SFR \citep{ellison08}.

\begin{figure*}
\epsscale{0.9} \plotone{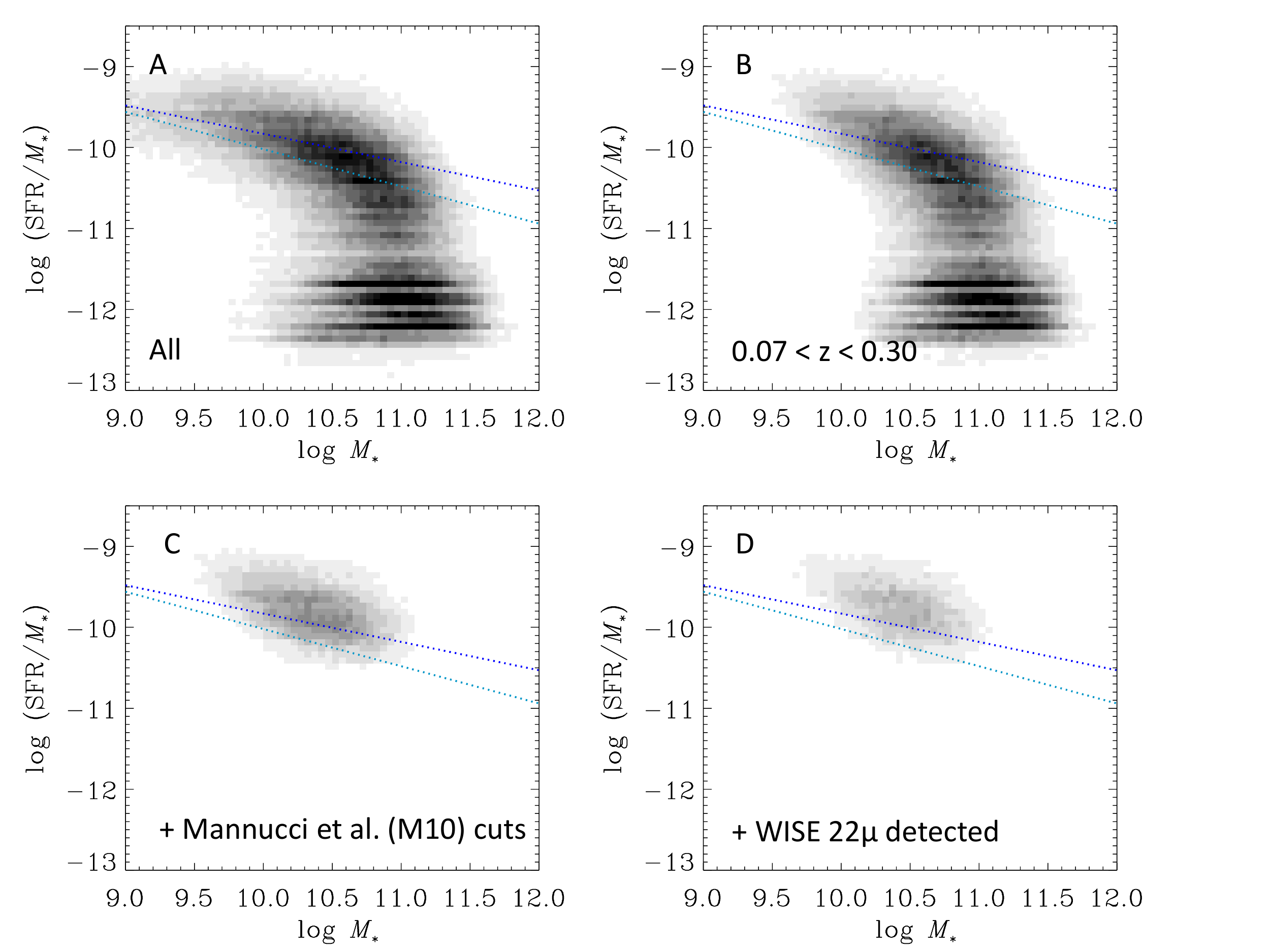}
\caption{Distribution of specific star formation rates (SFR/$M_*$) and
  stellar masses ($M_*$) illustrating the effects of sample cuts and
  detection limits. Panel A shows the sample (galaxies in the MPA/JHU
  SDSS DR7 catalog) prior to any cuts. Panel B shows the effects of
  applying the \citet{mannucci10} (M10) redshift cuts.  Panel C shows
  our final ``M10-like sample," where we have applied cuts that select
  star-forming galaxies with strong $\ha$ detection, and remove
  discrepant metallicities, following M10. In our analysis we will
  also use an {\it augmented} sample (plot not shown), which
  practically eliminates low-redshift limit as long as at least 10\%
  of mass is contained in the fiber. Panel D shows galaxies from the
  final M10-like sample (panel C) that have detections in \wise\ W4
  band ($22\mic$). A given shade of greyscale represents the same
  number of galaxies in every panel. Dotted lines show the location of
  the star-forming sequence according to \citet{s07} ($\beta=-0.35$,
  upper line) and \citet{sl12} ($\beta=-0.46$, lower line). SSFRs used
  in this plot come from the UV/optical SED fitting (but the sample is
  still only optically selected, see Section 2.3), so no biases in the
  SSFR--$M_*$ plane are introduced.\label{fig:ssfr_mass}}
\end{figure*}

To achieve a solid understanding of the process of chemical enrichment
and be able to interpret theoretical predictions, one first needs to
understand the root causes of the discrepant results
mentioned. Therefore, this paper will explore the uncertainties
regarding the character and the existence of the dependence of
metallicity on secondary parameters resulting from different methods
of measuring metallicity and SFR. Metallicity determinations have a
number of well-known difficulties-- both practical and theoretical
(e.g., \citealt{kewley08,andrews13}), so it is only natural that may
affect the characterization of the \mzs\ relation. Additional
systematics may arise from the fact that SDSS is limited to the use of
fiber spectroscopy, which samples only the central regions of galaxies
\citep{sanchez13}. Star formation indicators are similarly subject to
a number of caveats, including uncertainties in calibrations, the
types of populations used as tracers, and the dust corrections (e.g.,
\citealt{b04,s07,s09,lee09,lee11,kennicutt12,calzetti13}). Furthermore,
all of the previous work in SDSS used SFR estimates that are based on
some of the same emission lines measurements that are used to derive
metallicities, raising concerns about spurious correlations
\citep{lilly13}. The paper will also address possible biases arising
from sample definitions (e.g., \citealt{mithi}), and will revisit the
questions of the metallicity scatter and of the secondary dependence
of metallicity on galaxy size.

The present study will focus on characterizing the \mzs\ relation and,
more specifically, on verifying the existence of the SFR dependence
((i.e., $Z$($M_*$,SFR)) in the {\it local universe} ($z\sim0.1$). To
accomplish these goals we introduce an intuitive and physically
motivated framework which does not pre-suppose a particular
parameterization of this relation. The proposed framework can also be
naturally applied to establish whether the \mzs\ relation is
epoch-invariant, i.e., whether FMR is indeed fundamental. Note that
the question of the existence of FMR (the constancy of \mzs\ relation)
is separate from the question of the existence of $Z$($M_*$,SFR) at
any given redshift \citep{maier14}. Both of these questions, the
FMR and $Z$($M_*$,SFR) at $z\sim 2.3$, will be addressed in a
subsequent work (Salim et al.\ 2015, in prep.)

Throughout the paper we will use two-dimensional representations of
familiar quantities, as those are much more readily comprehended than
the three-dimensional representations. Striving to make the analysis
as intuitive as possible, we will be presenting only two types of
plots: mass vs.\ metallicity, and metallicity vs.\ the relative
specific SFR (for galaxies of a certain mass).

In Section 2 we present the data used in the study, define the
samples, and detail how different measurements were made. In Section 3
we motivate and lay out the non-parametric analysis framework that we
will use in the rest of the paper, and explore the character of
$Z$($M_*$,SFR) using a variety of SFR and metallicity measurements. In
Section 4 we apply our methodology in order to address some of the
discrepant results reported in the literature, while Section 5
discusses the implications for theoretical efforts. For a reader
interested in a quick overview of the results and their implications,
we suggest reading Section 6 first, followed by figure captions and
Section 4. Throughout the work we assume standard cosmology ($H_0=70$
km s$^{-1}$ Mpc$^{-1}$, $\Omega_m=0.3$, $\Omega_{\Lambda}=0.7$).

\section{Data, samples and measurements} \label{sec:data}

SDSS spectroscopic survey \citep{sdsssp} represents the largest survey
of galaxies in the local universe ($z<0.2$) and has served as a
primary source of data for studies that have described the MZR (T04)
and the \mzs\ relation (M10, LL10). Thus, in this paper we also use
SDSS as the basis for our work.

\subsection{Sample cuts}\label{ssec:sample}

In this section we describe the cuts we use to define the samples
drawn from SDSS spectroscopic survey, and illustrate their effect on
the properties of the sample in Figure \ref{fig:ssfr_mass}. Definition
of the sample in this paper follows the procedures adopted by M10 to
select the star-forming galaxies. As in M10, we start with the MPA/JHU
reduction of SDSS DR7 spectroscopic sample. MPA/JHU
catalog\footnote{\url{http://www.mpa-garching.mpg.de/SDSS/DR7}}
contains spectroscopic line measurements derived using a custom
pipeline that yields continuum subtracted fluxes (T04). Full MPA/JHU
SDSS DR7 sample consists of 928,000 galaxies (all sample numbers are
rounded to the nearest thousand), but this number drops to 212,000
after the application of the M10 redshift ($0.07<z<0.30$) and $\ha$
signal-to-noise ratio cuts (S/N($\ha$)$>25$). Relatively high
signal-to-noise cut on $\ha$ flux was applied in order to yield usable
signal in other, weaker, emission lines that are necessary for the
measurement of metallicities, while the relatively high low-redshift
limit was chosen to ensure that spectroscopic fibers cover large
fraction of each galaxy (2 kpc at $z=0.07$ and 7 kpc at
$z=0.30$). Line flux errors are taken as listed in the MPA/JHU
catalog.
 
Next we follow M10 and apply minor cuts to remove anomalously low
($F(\ha)/F(\hb)<2.5$) and very high Balmer decrements, i.e., dust
attenuations ($A_V>2.5$), bringing the sample to 203,000. The purpose
of these cuts is to eliminate unphysical or extreme dust attenuation
corrections. We presumed that M10 followed \citet{nagao06}, who
determine the attenuation at $V$ band from the Balmer decrement using
\citet{cardelli89} extinction law:

$$
A_V = 7.23\log\frac{F(\ha)/F(\hb)}{2.86},
$$

\noindent with the attenuation affecting the $\ha$ line being related
to $A_V$ by:

$$
A_{\ha} = 0.818 A_V.
$$

\noindent Note that $A_V$ is evaluated in $V$ band as a matter of
convention, but it represents extinction in $HII$ regions, which is
typically several times larger than the extinction of the stellar
continuum \citep{calzetti00,cf00}. Galaxies where emission line flux
is dominated by non-stellar emission are removed using the
\citet{k03c} criterion on the BPT diagram \citep{bpt} line ratios,
coupled with the log N2 $<-0.2$ cut (N2 = $F$([NII]6584)/$F(\ha)$),
leaving 160,000 galaxies.

In Figure \ref{fig:ssfr_mass} we show the effect of the cuts as
reflected in the specific SFR vs.\ stellar mass plane. Specific SFRs
are derived from the UV-optical SED fitting and masses are taken from
MPA/JHU catalog \footnote{Full description of various types of
  measurements is given in Section 2.3. The choice of SFR indicator in
  Figure \ref{fig:ssfr_mass} is of little importance.}. Figure
\ref{fig:ssfr_mass}A shows the initial sample. We see the well-known
bimodality in star-formation properties, with mostly quiescent
galaxies lying below log (SFR/$M_*) = -11.5$, and actively
star-forming galaxies lying on a relatively narrow sequence (the
'star-forming sequence' or the 'main sequence'). The upper dotted line
(repeated in all panels) represents the star-forming sequence fit from
\citet{s07} (hereafter S07), and pertains to galaxies selected as
star-forming according to \citet{b04} (hereafter B04) criteria, and
the lower line shows the SF sequence as determined in \citet{sl12}
using the gaussian decomposition of the full sample into star-forming
and passive galaxies. Application of the redshift cut (Figure
\ref{fig:ssfr_mass}B), applied following M10, removes the low-mass
tail, especially among galaxies with less intense SF. As a result, for
$\log\ M_*<10$ the sample tends to retain galaxies above the mean SF
sequence relation. Application of the remaining M10 selection
criteria, most importantly the $\ha$\ SNR cut and the BPT selection,
removes quiescent galaxies, but also eliminates a significant fraction
of galaxies on the SF sequence (Figure \ref{fig:ssfr_mass}C).

In consideration of the potential biases described above, we have also
constructed an {\it augmented} sample, which retains all the cuts as
the M10 sample, except that it allows redshifts extending down to
0.005, as long as the fiber contains at least 10\% (following
\citealt{yates12}) of the total stellar mass (fiber masses are
available from the MPA/JHU catalog). Even at lowest redshift the
majority of galaxies pass the fiber mass fraction criterion, so we do
not expect a bias due to incompleteness. Note that SDSS mass covering
fraction is typically 30\% (95\% of galaxies have the covering
fraction between 17\% and 50\%).  The augmented sample facilitates
fuller characterization of the trends at $\log M_*<10$ (plot not
shown). The augmented sample contains 259,000 galaxies (a 60\%
increase).

\subsection{Metallicities}

In this work we will use four metallicity indicators: (1) M10
metallicities---an average of R23 and N2 estimates, (2) Bayesian
metallicities of T04, (3) N2O2 and (4) R23 with an O32 term. We will
primarily use metallicity estimates determined following a procedure
described in M10. We refer to these metallicities as M10
metallicities. M10 metallicities are the average of \oh\ estimates
from two strong-line methods: R23 (the ratio of four oxygen lines
([OII]3727 doublet and [OIII]4958, 5007) to $\hb$, \citealt{pagel79})
and N2 (the ratio of [NII]6584 to $\ha$, \citealt{vanzee97}). R23 and
N2 metallicities were derived using relations from \citet{maiolino08},
which calibrate various line ratios with respect to \citet{kewley02}
theoretical metallicities of SDSS galaxies. All emission lines were
corrected for dust attenuation using $\ha$-to-$\hb$ Balmer decrement
and \citet{cardelli89} extinction curve. Following M10 we allow
galaxies with log R23 $<0.9$ and log N2 $<-0.35$ (the range of
validity of \citealt{maiolino08} calibrations), and remove those where
the two metallicity estimates based on R23 and N2 differ by more than
0.25 dex. We confirm that neither of these cuts removes galaxies of
any particular SFR range. The final M10-like sample (i.e., sample with
M10 redshift cuts) consists of 141,000 galaxies (matching the number
quoted in M10), while the final augmented sample (sample allowing
$z<0.07$) consists of 222,000 galaxies. The median redshift of the
final M10-like sample is 0.11, and of the final augmented sample is
0.08.

For comparison with previous work, and to discuss potential biases
affecting M10 metallicities, we will consider additional metallicity
estimates. One of them is the metallicity estimate taken from the
MPA/JHU catalog, which was derived following the methodology described
in T04. T04 metallicities are different from most other estimates
because they are not based on some line ratio calibrated against
another metallicity estimate, but come directly from fitting the
stellar-continuum subtracted spectra to emission-line photoionization
model spectra containing multiple emission lines (four Balmer lines
and eight forbidden lines). T04 metallicities are not publicly
available for all galaxies for which we calculate M10 metallicities
because MPA/JHU catalog provides them only for galaxies classified as
star-forming by B04 criteria. Those criteria required a S/N threshold
of 3 in all four BPT lines, but with flux errors scaled up to account
for the scatter in fluxes of repeat observations, thus effectively
becoming equivalent to S/N ratio cuts of between 5 and 7. As a result,
of 141,000 (222,000) galaxies in the final M10-like (augmented) sample
93,000 (163,000) have T04 metallicities ($\sim60$\%).

Recently, \citet{juneau14} revisited the analysis of repeat
observations and showed that flux error scalings are much smaller for
{\it flux ratios} (needed for BPT classification and metallicity
determinations) than for absolute fluxes, and are very close to
one. Therefore, in this paper we do not scale flux errors.

We will also derive metallicities using the N2O2 method (the ratio of
[NII]6584 to [OII]3727) for which theoretical calibration is taken
from Eq.\ 7 in \citet{kewley02}, and using the R23+O32 method (R23
method with an O32 ([OIII]4958, 5007 to [OII]3727
line ratio) term, \citealt{m91}), based on theoretical calibration from Eq.\
18 in \citet{kk04}. Unlike the M10 metallicities, these two methods
are expected either to be less dependent on the ionization parameter
(N2O2), or to explicitly correct for it (R23+O32).

\subsection{Star formation rates and stellar masses}

In this study we will primarily use four measurements of SFR. Two
pertain to SFR contained within the SDSS spectroscopic fiber, and two
are integrated (total) SFRs. Following M10 we calculate fiber SFR from
the $\ha$ luminosity, dust corrected with a Balmer decrement, and
converted to SFR using \citet{kennicutt98} conversion. Another fiber
SFR estimate comes from the MPA/JHU catalog, determined according to
B04 method of fitting photoionization models to six emission lines
(including $\ha$ and $\hb$) simultaneously. As in the case of T04
metallicities, the method does not apply any explicit conversions to
obtain the target parameter (metallicity or SFR), but performs
Bayesian fitting of fluxes or flux ratios. B04 have shown that
photoionization models imply that the conversion between the
dust-corrected $\ha$ and SFR is strongly metallicity (and therefore,
indirectly, mass) dependent. As a result, B04 SFRs can be as much
as 0.5 dex higher for the most metal-rich galaxies. Fiber SFRs,
originating from nebular emission of very massive stars, measure
instantaneous SFR (timescale $<10$ Myr).

First of the two total SFRs comes from the SED fitting of \galex\
\citep{martin05} ultraviolet (UV) and SDSS optical broad-band
photometry. We perform SED fitting on all SDSS spectroscopic sample
galaxies covered by \galex\ medium-deep imaging (50\% of SDSS
area). Fitting includes galaxies covered but not detected by \galex,
invariably non-SF galaxies, thus keeping the sample optically
selected. SED fitting is performed using Bayesian approach, by
comparing the observed fluxes with a library of model fluxes obtained
from stellar population synthesis models \citep{bc03} using some
priors on star-formation history, and attenuated according to the
\citet{cf00} dust extinction model. The methodology is described in
detail in S07. Principal differences with respect to S07 is that we
now use model priors optimized for star-forming galaxies (as described
in \citealt{dacunha08}) and we produce model libraries more finely
sampled in redshift (0.01 vs.\ 0.05 in S07). SED SFRs, being mainly
constrained by UV, are determined as averaged over the past 100 Myr.

\begin{figure*}
\epsscale{1.15} \plotone{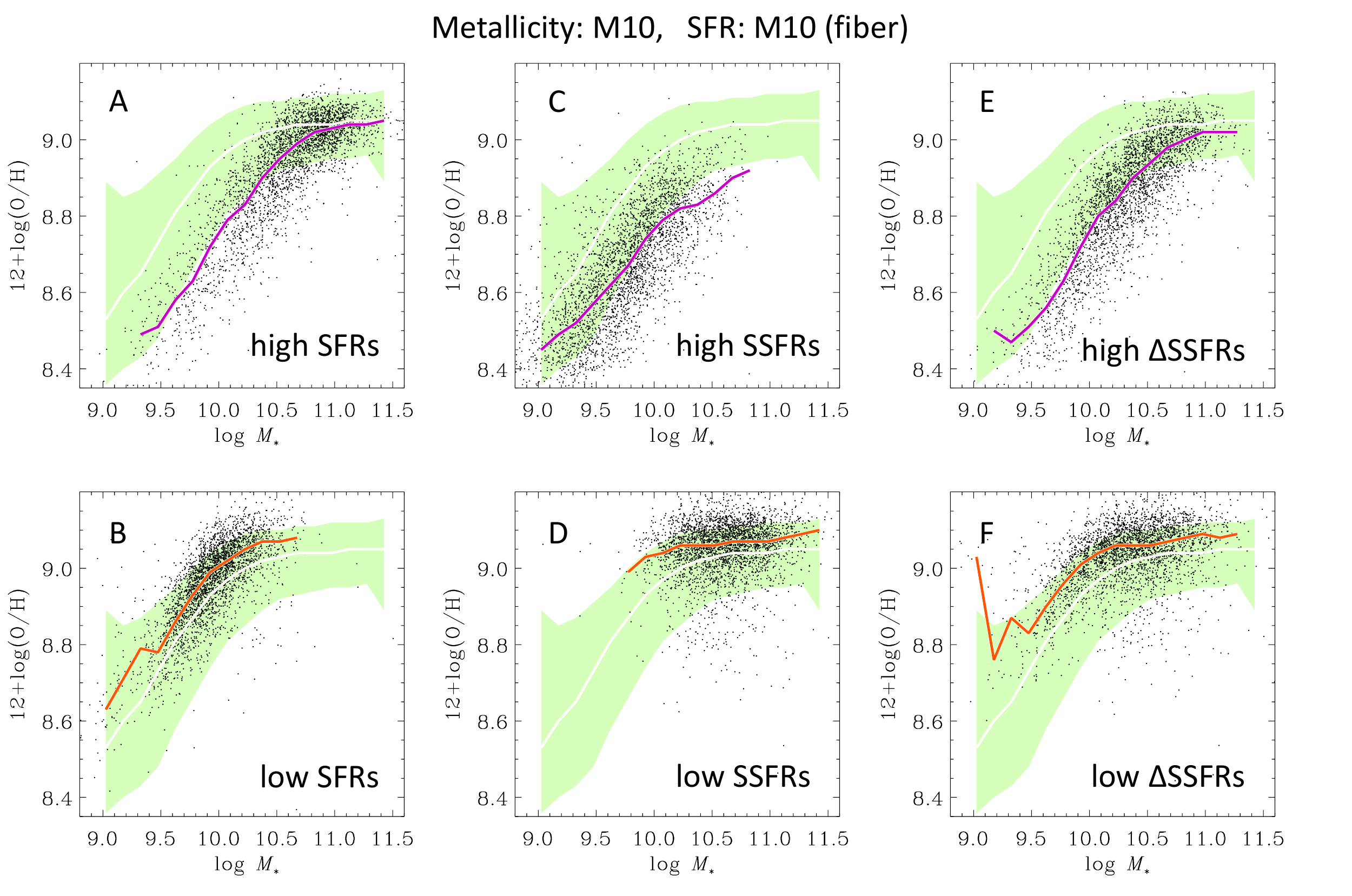}
\caption{Mass-metallicity diagrams of galaxies selected to be extreme
  according to some SFR-related parameter.  The upper (lower) row of
  panels show galaxies with the highest (lowest) 2.5\% of values in
  the sample (for E and F, extreme in each mass bin).  Sample
  selection follows M10 (our ``M10-like" sample), which features an
  $\ha$ SNR cut of 25, but no SNR cuts on other emission lines. Panels
  A and B select by SFR, C and D by specific SFR, and E and F by
  specific SFR in a given mass bin. Fiber-based measurements are used
  in all panels. In all panels, green shaded regions give the 90
  percentile range of the overall M10-like sample, with the white line
  representing the median. All mass bins are 0.15 dex wide. Medians of
  the high/low samples (colored lines) are showed when more than 5
  galaxies exist in a bin.\label{fig:mz_resid_type}}
\end{figure*}

\begin{figure*}
\epsscale{1.15} \plotone{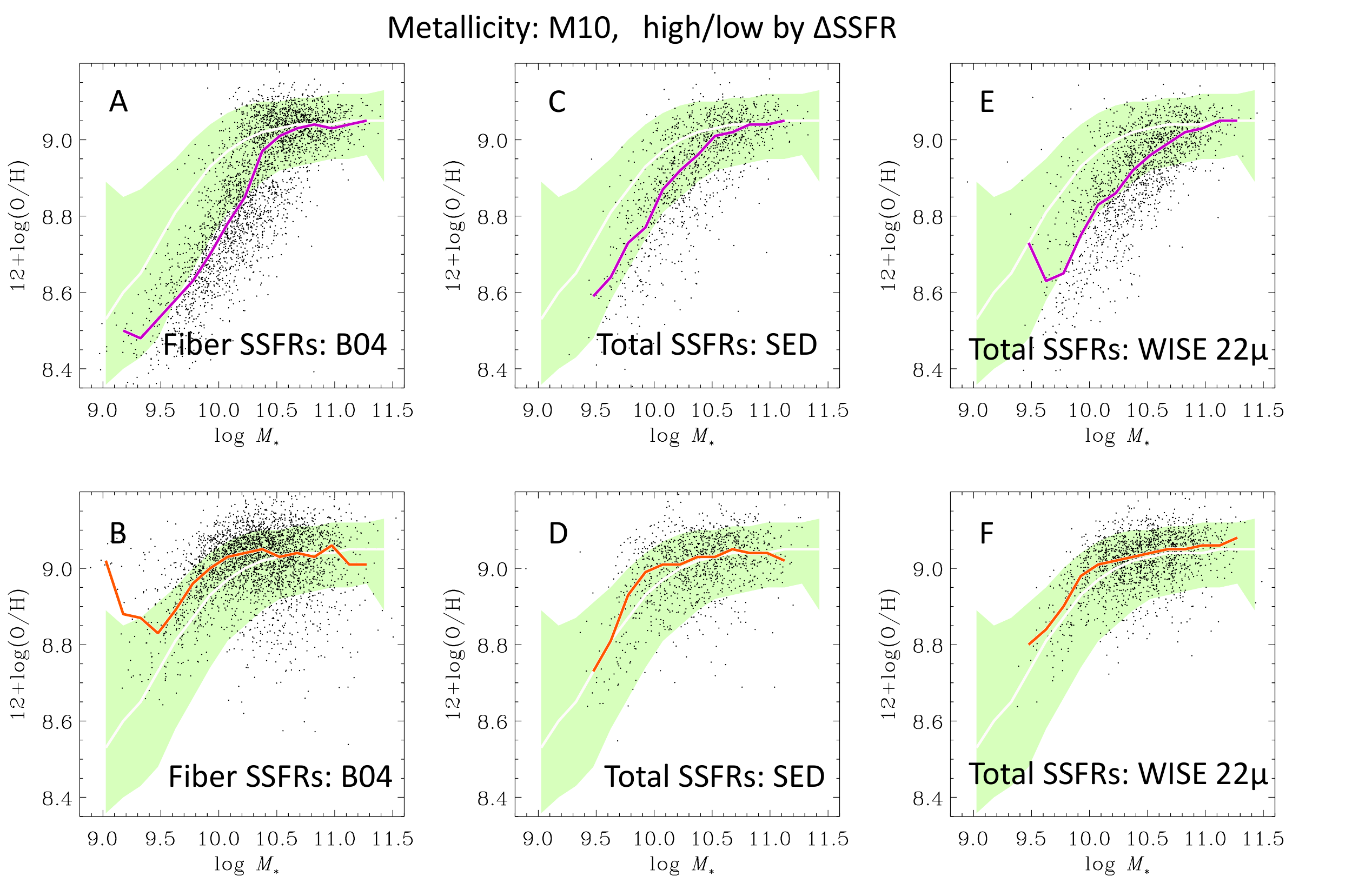}
\caption{Mass-metallicity diagrams of galaxies selected to have
  extreme SSFRs at fixed mass, where the SFR has been calculated using
  different indicators.  As in the previous figure, the upper (lower)
  row of panels show galaxies with the highest (lowest) 2.5\% of
  values in the sample, and the green region shows the 90 percentile
  range of the overall ``M10-like" sample. Panels A and B show samples
  selected by \citet{b04} fiber SSFR, C and D by total SSFR derived
  from broad-band UV/optical SED fitting, and E and F by total SSFR
  derived from $22\mic$ mid-IR luminosity (\wise\ channel
  W4).  \label{fig:mz_resid_sfr}}
\end{figure*}

Second type of total SFRs come from 22$\mic$ (W4 channel) observations
from \wise\ \citep{wise}. We use the AllWISE catalog profile-fit W4
fluxes\footnote{We confirmed that mid-IR SFRs from the profile-fit
  photometry produce smaller scatter with respect to other SFRs than
  the standard aperture \wise\ photometry.} and extrapolate them using
\citet{dh02} IR SEDs calibrated with \citet{marcillac} relations to
obtain the total IR luminosity, which is then converted to SFR using
the \citet{kennicutt98} relation. Details can be found in \citet{s09},
where the same procedures were used on {\it Spitzer} 24$\mic$
fluxes. Detections at 22$\mic$ are available for 54\% of the final
sample, but this incompleteness does not produce strong biases in
SSFR--$M_*$ parameter space (Figure \ref{fig:ssfr_mass}D). Mid-IR
emission is typically assumed to originate from dust-enshrouded young
populations that also give rise to the UV emission and therefore are
expected to trace current ($\sim$ 100 Myr) SFR. However, \citet{s09}
have shown that mid-IR may have significant contributions from older
populations even in actively star-forming galaxies, so that the
timescale over which mid-IR measures SF may be effectively on the
order of few Gyr.

We will also be considering specific SFRs. Fiber SSFRs are obtained by
normalizing fiber SFRs by fiber stellar masses from the JHU/MPA
catalog. SED SSFRs are derived directly from the SED fitting and
mid-IR SSFRs are normalized by the total stellar mass from the JHU/MPA
catalog. As in M10, we use stellar masses from the the JHU/MPA
catalog, which were derived from SDSS photometry (following, but
independently from S07). These masses are in a very good agreement
(scatter in difference $\sim0.08$ dex, with $<0.01$ dex overall
offset) with the masses obtained from our SED fitting.

For all MPA/JHU measurements (metallicity, SFRs, stellar masses) and
for parameters from the SED fitting we use medians of the probability
distribution function as fiducial parameter values.

SFRs and stellar masses were either derived with or converted to
Chabrier IMF.

\section{\mzs\ relation in the local universe: empirical analysis framework and systematics}

%
\subsection{Setting the stage: identifying the physically motivated
  secondary parameter in the $M_*-Z$ relation}

In this section we set the stage for the remainder of our analysis by
identifying the optimal and the most physically motivated parameter
driving the secondary dependencies in the MZR. We then use this
knowledge to propose an intuitive framework for investigating the
\mzs\ relation in Section 3.3.

A useful way to think of the \mzs\ relation is that it is represents
an extension of a familiar MZR. Thus, a commonly used way to
illustrate the SFR dependence is to show average (or median)
mass-metallicity tracks of galaxies selected to lie in different bins
of {\it absolute} SFR (e.g., M10,
\citealt{yates12,andrews13,nakajima14}). In this section we will show
that the absolute SFR is not the optimal quantity with which to
characterize MZR residuals. The choice to use absolute SFR as a
secondary parameter in the mass-metallicity relationship is probably
motivated by the fact that the \mzs\ relation has originally been
formulated by M10 and LL10 as the relation between the mass,
metallicity and SFR. Instead, we argue that this should rather be the
{\it relative} SSFR---the difference between galaxy's SSFR and the
SSFR typical for galaxies of that mass (i.e., the offset from the
star-forming sequence).

To show this, we will contrast MZRs (Figure \ref{fig:mz_resid_type})
of samples selected to have the extreme (highest 2.5\% (top panels))
and the lowest 2.5\% (lower panels)) values of (i) absolute SFR (left
panels), (ii) absolute SSFR (middle panel) and (iii) relative SSFR
(right panels) to the MZR of the general population of galaxies
(repeated green band in each panel is the 90 percentile range of the
metallicities of the overall M10-like sample). Median trends of top
(bottom) samples is shown as a purple (red) line, and of the general
population as a white line. In this section (3.1) we will only be using
M10 metallicities and SFRs derived in SDSS spectroscopic fibers, as
derived in M10. Thus our Figure \ref{fig:mz_resid_type} is directly
comparable to Figure 1 in M10.

We begin this exercise by exploring MZR offsets due to absolute SFRs.
We see that the galaxies with the highest SFRs (Figure
\ref{fig:mz_resid_type}A) show no offset with respect to the overall
sample at the highest masses ($\log M_*>11$), while at $\log M_*=10$
the offsets have increased to 0.2 dex, and stay that large at lower
masses. Since SFRs on average increase with mass (e.g., B04), the
highest SFRs are preferentially found among the more massive
galaxies. Likewise, the galaxies with the lowest SFRs will be found in
the region of lower masses (panel B). Galaxies with the lowest SFRs
are systematically offset {\it above} the median overall MZR, by some
0.05 dex, regardless of the mass.

Given that the SFR to first order simply scales with mass, it is
justified, and physically more motivated (e.g.,
\citealt{ellison08,yates12}), to explore whether the SFR normalized by
mass, i.e., the specific SFR--an indication of the current SF activity
with respect to the past average--would produce stronger offsets
across the mass range than just the SFR. We again show $M_*-Z$ plots,
but now for 2.5\% of the sample with highest and lowest SSFRs in the
sample (Figure \ref{fig:mz_resid_type}C and D). Since on average the
specific SFR declines with increasing mass (e.g., S07, also Figure
\ref{fig:ssfr_mass}), we now have the reverse situation that the
highest SSFRs are found among the lower-mass galaxies and the lowest
SSFRs are among the more massive. Median offsets for the intense
star-formers remain large at high masses, but they are somewhat
smaller than in the case of top SFRs for the lowest masses. For
galaxies with low SSFRs (panel D) there is not much overlap in the
mass regime with low SFRs (panel B), and the median MZR also sits some
0.05 dex above the overall median.

One can see that considering galaxies selected by either SFRs or SSFRs
implicitly introduces a mass selection. A true secondary parameter
driving offsets in MZR should apply to galaxies of all masses. Thus we
now explore samples selected by the level of SSFR {\it relative} to
what is typical at a given mass. Relative SSFR ($\Delta \log {\rm
  SSFR}$) is defined as:

$$
\Delta \log {\rm SSFR} = \log {\rm SSFR} - \langle\log {\rm SSFR}\rangle_{M_*},
$$

\noindent where $\langle\log {\rm SSFR}\rangle_{M_*}$ is the median or
mean of $\log {\rm SSFR}$ of galaxies having a mass $M_*$. In this
work we use medians in 0.15 dex wide mass bins. Relative SSFR can be
visualized as the offset of a galaxy from the star-forming sequence
(vertical distance in Figure 1), a measure also called ``SFR excess''
\citep{schiminovich07} or ``starburstiness'' \citep{elbaz11}. The
relative SSFR is a ``natural'' parameter to characterize
star-formation activity, and is becoming increasingly used in the
recent literature (e.g.,\citealt{magdis12,woo13}). In Figure
\ref{fig:mz_resid_type}E and F we now show galaxies with the highest
and lowest 2.5\% relative SSFRs in each mass bin. These galaxies
typically have SSFRS that are five times higher/lower than typical
values at that mass $|\Delta \log {\rm SSFR}|>0.7$ dex). The quantity
of points at different masses now reflects the mass distribution in
the overall sample. We notice that some offset for highly star-forming
galaxies (panel E) is now present even at the highest masses, which
was not the case when absolute SFRs were considered (panel A). At
lower masses the offsets are at least as strong as they were in panel
A, but now the selection includes more galaxies. Similarly, for
galaxies with the lowest relative SSFRs (panel F), the offsets are as
large as in the case of either the lowest SFRs or SSFRs, but spanning
the full range of masses.

From this section we conclude that MZR offsets are more naturally
characterized by the difference between the logarithm of galaxy's SSFR
with respect to a typical log SSFR at a given mass, rather than the
absolute SFRs. In hindsight this may seem fairly obvious, but this
point has not has not previously been clearly made. The interpretation
of the MZR offsets in terms of the variations of relative SSFR is more
in accordance with FMR's evolutionary sense: at higher redshifts the
SF sequence appears to shift upwards without much change in the slope
(e.g., \citealt{speagle}), so considering local samples at some
distance (offset) from the SF sequence is more analogous to a
high-redshift selection.

Does the above mean that the functional form of $Z$($M_*$,SFR) should
feature SSFR rather than SFR? In the case of a linear relationship and
{\it total} measurements, the two formulations are formally
equivalent. However, for {\it fiber} measurements, the difference is
crucial. Fiber SFR, used in M10's formulation of FMR (their Eqn.\ 2, 4
and 5), is not a meaningful physical quantity since it depends, to the
zeroth order, on galaxy distance: on average SDSS fibers cover 25\% of
SF at $z=0.07$, but 65\% at $z=0.30$. On the other hand, fiber {\it
  specific} SFR (fiber SFR normalized by mass in the fiber) is a
perfectly valid physical quantity, representing the intensity of SF in
the {\it same physical region} of the galaxy in which the metallicity
is measured, and is therefore probably even preferred to the total
specific SFR for the purposes of \mzs\ analysis. Future studies should
report their fiber or slit SSFRs to allow for a more direct comparison
with SDSS. Surprisingly, M10 find that their least-scatter projection,
formulated with {\it fiber SFRs}, agrees with high-redshift
measurements, even though fiber SFRs are distance-dependent and are on
average 0.6 dex smaller than the total SFRs (but with a substantial,
distance-dependent scatter).  Altogether, it makes more sense to
formally describe \mzs\ relations using specific SFRs. Indeed, the
recent analytical model of \citet{lilly13} (e.g., their Eq.\ 40) finds
the metallicity to be the function the SSFR (see discussion in Section
5).

Guided by the inferences made in this section, in the rest of the
paper we will consider the relative SSFR as the primary independent
variable for galaxies of a fixed mass.

\subsection{How dependent is \mzs\ relation with respect to the
  type of SFR indicator?}

Previous work on\mzs\ relation in SDSS used exclusively SFRs derived based on the emission
lines measured in spectroscopic fibers. As some of those same line measurements
are involved in the determination of metallicity, this open up a concern
that the \mzs\ relation may to some extent be the result of spurious correlations
between the measurements of SFR and the metallicity. Therefore, in
this paper we will examine the relationship with two completely independent
total SFRs, based on integrated fluxes. Furthermore, different SFR
indicators are sensitive to SF over different timescales, which in
principle may be more or less strongly tied to the changes in the metallicity. 

In this section we investigate four SFR indicators: two measured in
fibers (both based on emission lines, but one following M10's common
methodology of Balmer decrement-corrected $\ha$ luminosity, and the other
using B04's more sophisticated methodology of modeling
simultaneously multiple emission lines) and two total measurements
(one based on the UV/optical SED fitting, and the other based on
22$\mu$ mid-IR luminosity from {\it WISE}). The results are presented
in Figure \ref{fig:mz_resid_type} (right panels--M10 SFRs), and are
continued in Figure \ref{fig:mz_resid_sfr} (left panels--B04 SFRs;
middle panels--SED SFRs; right panels--mid-IR SFRs). In Figure
\ref{fig:mz_resid_sfr} we continue to plot the galaxies with 2.5\%
highest/lowest {\it relative} SSFRs (top/bottom panels), and contrast
them to the general population of galaxies (green band).  We continue to
use M10 metallicities.

We begin with the comparison of $M_*-Z$ plots based on M10 fiber SFRs
(Figure \ref{fig:mz_resid_type}E and F) to B04 fiber SFR (Figure
\ref{fig:mz_resid_sfr}A and B). Below $\log M_*=10.3$ there is not
much difference in MZRs of either the high or the low SSFR samples
selected by either M10 or B04 SSFRs. Above $\log M_*=10.5$, galaxies
selected by B04 SSFR show no offset with respect to the overall
metallicity. Since B04 SFRs are more sophisticated than M10 SFRs and
therefore presumably more accurate, this suggests that there is very
little or no SFR dependence in the MZR above $\log M_*=10.5$.

Next we look at SED SFRs---total SFRs determined from the UV/optical
broad-band fluxes, and primarily constrained by the UV. The MZR of the
most intensely star-forming galaxies (Figure \ref{fig:mz_resid_sfr}C)
shows approximately two times smaller offset below $\log M_*=10.3$
with respect to equivalent relations based on either fiber
SFRs. Above, $\log M_*=10.5$, the offset is nearly gone, as in the
case of B04 fiber SFRs. For galaxies selected to have the lowest
relative SSFRs (Figure \ref{fig:mz_resid_sfr}D) the offset is likewise
smaller than for fiber SFRs and basically not present above $\log
M_*=10.3$, again similar to B04 SFRs. Before drawing any conclusions,
we examine high/low samples selected by mid-IR SFRs. For intense
star-formers (Figure \ref{fig:mz_resid_sfr}E), the offset in MZR is,
surprisingly, again as strong as it was for M10 fiber SFRs, and is
clearly present even at $\log M_*=10.5$, which was not the case for
either B04 fiber SFRs or for the total SED SFRs. For galaxies furthest
below the SF sequence (but detected at 22$\mic$) the offset is quite
small (0.02 dex), as in the case of SED SFRs, and unlike M10 SFRs. We
have verified that this small offset is not because of 22$\mic$
detection limit potentially eliminating lowest star-formers---if we
produce M10 SFR selected $M_*-Z$ plot (not shown) but only for
galaxies also detected at 22$\mic$ we still obtain an offset as large
as the one in Figure \ref{fig:mz_resid_type}F.

To conclude, we find that the dependence of MZR on SSFR is definitely
present using all SFR indicators, whether they be fiber or total, so
the \mzs\ relation cannot be merely an artifact of correlated
measurements. The degree of offsets vary, but at least one SSFR of
both types (fiber and total) shows equally strong offsets for highly
SF galaxies. It is interesting that the dependence is not
significantly weakened when using total SSFRs, considering that the
metallicity is measured in the fiber. Note that if an SSFR measure has
a high uncertainty, it will not be able to accurately identify
galaxies that are truly the highest star-formers and therefore the
most discrepant in metallicity, which may explain why the trends are
weaker using SED SFRs. This emphasizes the need to perform comparison
of samples at different redshifts using indicators of similar
accuracy, and, preferably, of the same type. Results further suggest
that the fact that different indicators trace SF over different
timescales does not appear to affect the trends
significantly. Finally, we confirm previous findings
(\citealt{ellison08}, M10) that the MZR offsets are more pronounced at
lower masses, and that they are close to zero when approaching $\log
M_*=11$.

\begin{figure*}
\epsscale{1.1} \plotone{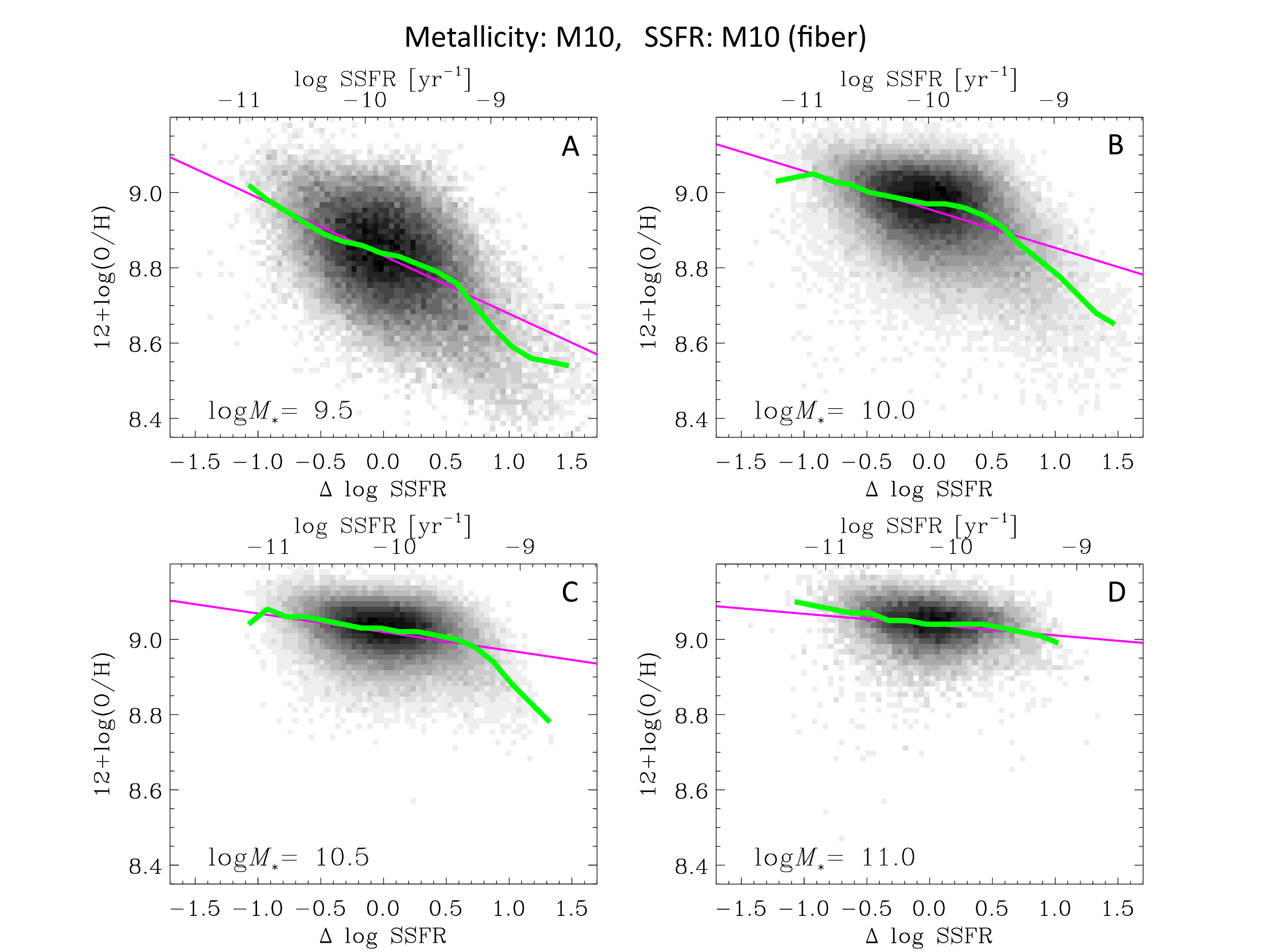}
\caption{Dependence of the metallicity on the offset from the
  star-forming sequence, in different mass bins. Both the metallicity
  and the fiber SFRs are as derived in M10. We now use an augmented
  sample, which follows the M10 selection (most importantly the $\ha$
  SNR cut of 25) but allows redshifts down to 0.005 as long as the
  mass contained in fiber is at least 10\% of the total galaxy
  mass. Mass bins are centered on the values indicated in each panel,
  and are 0.5 dex wide. The absolute SSFR at the center of the mass
  bin is given along the top of each plot.  Magenta lines show linear
  fits to the data points, while the green lines represent medians
  (when at least 15 galaxies exist in a 0.15 dex wide bin). Greyscale
  uses square-root scaling in order to better display the full dynamic
  range of the density of data points. A SSFR dependence is present at
  all masses, but is stronger at lower masses. It also depends on the
  SSFR itself, and is stronger above the star-forming sequence in the
  three lower mass bins. \label{fig:dssfr_z_m10}}
\end{figure*}

\subsection{Characterization of the \mzs\ relation}

In previous sections we used familiar $M_*$--$Z$ plots and contrasted
samples of galaxies selected to be extremes in terms of star-forming
properties. We now wish to include all of galaxies in the analysis. We
have also seen that the MZR offsets are strongly mass dependent. We
therefore need to analyze galaxies of different mass ranges
separately. Following the framework set up in Section 3.1, we do so by
plotting the metallicities against the relative SSFR. These
$Z$--$\Delta$SSFR plots for galaxies belonging to some mass bin embody
our non-parametric methodology for exploring the \mzs\ relation.

In Figure \ref{fig:dssfr_z_m10} we show four $Z$--$\Delta$SSFR plots,
for one of 0.5 dex wide mass bins centered on $\log M_* =9.5$, 10.0,
10.5 and 11.0. Results do not depend strongly on the width of mass
bins. We continue to use M10 metallicities and revert to M10 fiber
SSFRs, but from now on use the {\it augmented} sample (one extending
to lower redshift, $z=0.005$, than what was used in M10), which allows
for better characterization of the low-mass regime (Section
\ref{ssec:sample}). Relative SSFR for a given galaxy is determined as
the difference of its SSFR with respect to the median SSFR in 0.15 dex
wide mass bin. Positive relative SSFRs correspond to galaxies sitting
above the star-forming sequence. Almost identical results would be
obtained if, instead of using running medians, the relative SSFR was
calculated as the difference between SSFR and the SSFR corresponding
to the linear fit of log SSFR vs.  $\log\ M_*$ at that mass (such as
those in Figure 1). Figure \ref{fig:dssfr_z_m10} now allows us to
explore dependence of $Z$ on SSFR for all galaxies, not just
those at the extreme of SSFR distribution.

We start from the mass bin centered on $\log\ M_*=9.5$ (Figure
\ref{fig:dssfr_z_m10}A), where the metallicity trend is stronger than
in higher mass bins. Fitting the linear relationship (purple line)
yields a slope $\kappa = d(12+\log({\rm O/H}))/d\log ({\rm SFR}/M_*)$
of $-0.18$. However, running medians reveal that the SFR dependence is
considerably stronger above the SF sequence (a linear fit would yield,
$\kappa_{\rm high} = -0.26$), then in its core and below it
($\kappa_{\rm low} = -0.12$). Moving on to the next mass bin (Figure
\ref{fig:dssfr_z_m10}B), we find that the overall dependence on SFR
gets weaker ($\kappa=-0.12$), but again the slope is steeper above the
SF sequence than in its core and below it. Remarkably, the slope above
the SF sequence is as steep as in the lower mass bin (values of slopes
are given in Table \ref{table:char}). At $\log M_*=10.5$ (Figure
\ref{fig:dssfr_z_m10}C), the overall slope is only $\kappa=-0.05$, and
while it is again steeper above the SF sequence (and still as steep as
at lower masses), the steepening does not begin until 0.7 dex above
the sequence, and consequently encompasses only a small number of
highly star-forming galaxies. Finally, in $\log M_*=11.0$ bin (Figure
\ref{fig:dssfr_z_m10}D), we stop seeing different behavior above and
within/below the SF sequence, with a rather shallow overall trend of
$\kappa=-0.03$.

Are the results presented here valid for total SSFRs? We list the
slopes of metallicities against the offset from the SF
sequence based on mid-IR SFRs in Table \ref{table:char}). The results
are remarkably similar to those based on fiber SFRs, with the most
notable difference being that the slope in the highest mass bin is
even weaker. This basically confirms the conclusion from Section 3.2
that the differences in timescales of SFR indicators and whether they
pertain to fiber or integrated measurements do not lead to great
differences in metallicity's dependence on SSFR.

To conclude, in this section we have demonstrated that the
metallicity's dependence on SSFR is stronger {\it above} the SF
sequence (as remarked by M10). And while the dependence for the bulk
of galaxies (those in and below the SF sequence) weakens as the mass
goes up, the dependence for intense star-formers (lying at $\gtrsim
0.6$ dex from the SF sequence) stays very similar, suggesting that the
different mechanisms drive the \mzs\ relation depending on SF
intensity, which itself may be related to the existence of different
modes of SF (e.g., quiescent vs.\ merger-driven). Some studies,
especially at $z\gtrsim 1$, distinguish populations with high relative
SSFR (e.g., $\Delta$ log SSFR$>0.3$, \citealt{elbaz11}),) as having a
special mode of star formation, associated with mergers, and label
them ``starbursts''. However, at $z\sim2$, the starbursts produce a
clear excess above the Gaussian distribution of log SSFR and dominate
already at $\Delta$ log SSFR$=0.6$ \citep{rodighiero11}. However,
while we find the high-end distribution of log SSFR in SDSS to
eventually depart from a Gaussian, this excess does not become
dominant (twice as high as the Gaussian) until $\Delta$ log SSFR$>1.0$
dex (for M10 SFRs, at $\log M_*=10$), well above the onset of break in
the $Z$--SSFR relation. Therefore, we refrain from equating the
two-mode behavior in metallicity trends with the normal SF vs.\
merger-driven starburst distinction at this time, but do not rule out
such a connection.

\begin{deluxetable*}{l r r r r r r r r} 
 \tablecaption{Characterization of the \mzs\ relation. \label{table:char}}
\tiny
\tablewidth{0pt}
\tablenum{1}
 \tablehead{
   \colhead{SFR type}      &
   \colhead{$\log M_*$}      &
   \colhead{Median 12+log(O/H)} &
   \colhead{$\sigma$(\oh)} &
   \colhead{$\sigma_{\rm corr}$(\oh)} &
   \colhead{$\kappa$} &
   \colhead{$\kappa_{\rm low}$} &
   \colhead{$\kappa_{\rm high}$} &
   \colhead{Break point}\\
   }
\startdata
M10 (fiber) & 9.5 & 8.83 & 0.121 & 0.102 & $-0.15$ & $-0.12$ & $-0.27$
& 0.6\\
M10 (fiber) & 10.0 & 8.96 & 0.092 & 0.081 & $-0.10$ & $-0.06$ & $-0.29$
& 0.5\\
M10 (fiber) & 10.5 & 9.02 & 0.063 & 0.059 & $-0.05$ & $-0.04$ & $-0.32$
& 0.7\\
M10 (fiber) & 11.0 & 9.03 & 0.052 & 0.051 & $-0.03$ & $-0.03$ & $-0.16$
& 0.7\\
\hline 
WISE 22$\mic$ (total) & 9.5 & 8.85 & 0.119 & 0.107 & $-0.17$ & $-0.14$ & $-0.16$
& 0.3\\
WISE 22$\mic$ (total) & 10.0 & 8.96 & 0.089 & 0.082 & $-0.11$ & $-0.08$ & $-0.19$
& 0.5\\
WISE 22$\mic$ (total) & 10.5 & 9.02 & 0.060 & 0.058 & $-0.04$ & $-0.03$ & $-0.21$
& 0.6\\
WISE 22$\mic$ (total) & 11.0 & 9.04 & 0.049 & 0.049 & $-0.01$ & $-0.01$ & $-0.22$
& 0.7\\
\enddata
\tablenotetext{}{Median 12+log(O/H) is given at the position of the SF
  sequence according to the linear fit. Scatter in metallicities
  around the running median, i.e., corrected for SFR dependence is
  $\sigma_{\rm corr}$(\oh). Slope of the metallicity vs.\ the change
  in relative SSFR (i.e., the offset from the SF sequence) is $\kappa$, while
  the slopes above and below the break point (dex above the SF
  sequence) are $\kappa_{\rm high}$ and $\kappa_{\rm low}$. All
  metallicities are derived as in \citet{mannucci10} (M10).}
\normalsize
\end{deluxetable*}

Important implication of this finding is that describing or
extrapolating the SFR dependence using simple linear trends
(equivalent to assuming a flat ``fundamental plane'' in LL10 or a
single preferred projection of the FMR in M10) could lead to
inconsistencies when comparing to high-redshift samples, as the local
trends will be dominated by galaxies that show weaker SFR
dependence. We see that using a single linear trend can produce
discrepancies as large as 0.2 dex in the metallicity for the most
active star-formers. Such difference can give very different character
to the interpretation of high-redshift data. Recently \citet{maier14}
showed (their Figure 5) that using different descriptions of the \mzs\
relation by M10 (second order polynomial vs.\ the plane of the least
scatter) produce different extrapolations for high-SSFR samples. At a
given mass, a better way to parametrize the SSFR dependence is with a
broken linear fit---one for galaxies above the SF sequence, and other
for the rest. Ultimately, to test whether high-redshift samples follow
the local \mzs\ relation (and to determine whether they exhibit the
dependence on SSFR internally) it is best to show both the local and the
high-redshift data on metallicity vs.\ SSFR plots, within some mass
bin (lower masses will provide stronger diagnostic as long as samples
are not too small). We apply this methodology in Salim et al.\ (in
prep.) to test whether local \mzs\ relation is consistent with $z\sim
2.3$ measurements.

\subsection{Does accounting for SFR lead to a considerable decrease in
  the scatter of the MZR?}

The existence of the (S)SFR-dependence of MZR implies that if the
(S)SFR was accounted for, the scatter in the relation would decrease.
This decrease was presented in M10 as being dramatic. For example,
their Fig.\ 5 shows that the scatter in metallicity becomes a factor
of almost three smaller (goes from 0.055 dex to 0.02 dex). However,
what is perhaps not sufficiently appreciated is that this calculation
in M10 pertained to the reduction of the RMS residuals of {\it
  median-binned} values of metallicity around the best-fitting
surface, and not of the individual galaxies.  Different studies have
claimed conflicting results regarding the scatter of individual
galaxies (non-binned samples). \citet{ellison08} reported that SSFR is
not an important cause of scatter in MZR (a 10\% reduction), and more
recently \citet{perez-montero13} quoted a 0.01 dex reduction. These
results are at odds with M10 stated reduction from 0.08 dex to
0.05--0.06 dex.

The methodology applied in Section 3.3 allows us to directly address
the question of the reduction of scatter. Figure \ref{fig:dssfr_z_m10}
shows that the scatter in metallicity is high even at a fixed mass and
fixed SSFR (metallicity axis in Figure \ref{fig:dssfr_z_m10} spans the
exact same range as in $M_*-Z$ plots: Figures \ref{fig:mz_resid_type}
and \ref{fig:mz_resid_sfr}). Even at low masses ($\log M_*=9.5$),
where the SFR dependence is the strongest, the overall scatter
(standard deviation) in metallicities of 0.12 dex is reduced only to
0.10 dex (scatter around the median), a 20\% reduction. The effect is
even more modest at higher masses, where the dependence on SFR is
weaker. The values of the standard deviation of metallicities before
and after accounting for the SFR are given in Table
\ref{table:char}. The uncertainty in the measurement of SSFR ($\sim
0.2$ dex) adds to the scatter in $Z$, but this contribution ($0.2
\kappa$) does not account for more than 10\% of the residual
scatter. Based on this we conclude that presenting the MZR using a
quantity that combines the mass and SFR (as in M10 Fig.\ 5), should
not be expected to reveal conspicuous reduction in the scatter for
typical samples of galaxies. On the contrary, the reduction is indeed
very modest. This also means that the \mzs\ relation, as a three-dimensional
``surface'', is not particularly thin, despite common wisdom (e.g.,
\citealt{steidel14}), possibly inspired by M10's use of the {\it
  binned} points (e.g., their Figure 2).\footnote{To illustrate how
  binning can lead to a wrong impression, imagine a homogeneous
  spherical distribution of points in three-dimensional space
  $xyz$. Now perform binning of $z$ values in bins defined in the $xy$
  plane. The binned values would form a perfectly thin {\it plane}.}

Finally, we turn to another question regarding the scatter in $Z$. T04
and \citet{zahid12} have previously noted that the metallicity scatter
in MZR, which becomes quite large at lower masses, is not due to, for
example, larger measurement errors, but is intrinsic. The question is
whether it can be fully explained by dependence on SFR. As we have seen,
taking into account the SSFR dependence reduces the scatter at $\log
M_*=9.5$ from 0.12 only to 0.10 dex, which is still significantly
larger than the typical metallicity errors at that mass (0.04 dex,
T04). This suggests that some other parameter(s) may ultimately be
more closely connected with metallicity regulation than the SFR.

\begin{figure*}
\epsscale{1.1} \plotone{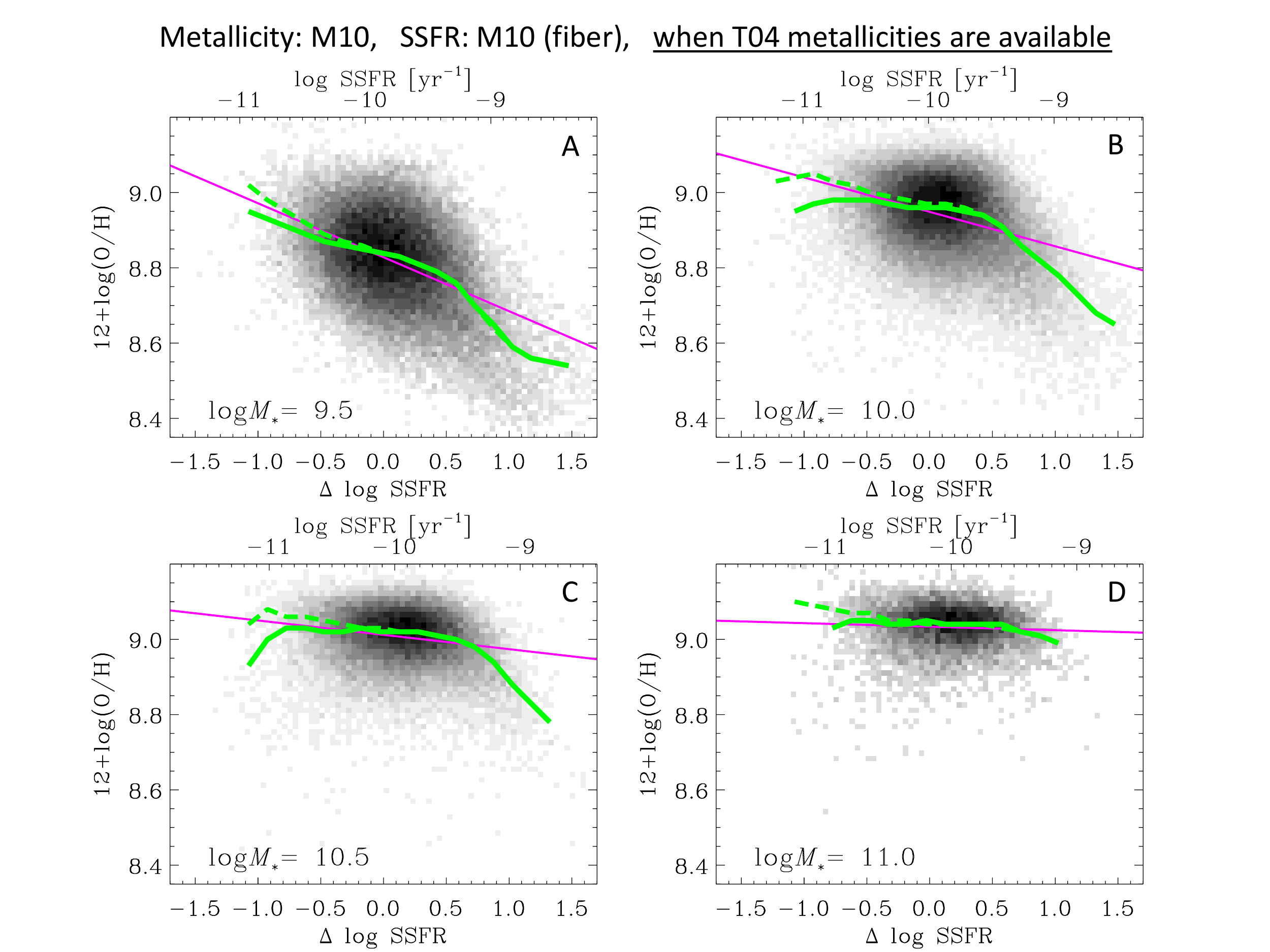}
\caption{Dependence of metallicity on the offset from the star-forming
  sequence in different mass bins. This figure uses the same
  measurements as Figure \ref{fig:dssfr_z_m10}, except that it shows
  only those galaxies from the augmented sample ($\sim$2/3) for which
  \citet{tremonti04} (T04) metallicities are available. Unlike the
  selection in M10 sample, T04 metallicities require relatively high
  ($>$5--7) signal-to-noise ratio in all four BPT lines. Such
  multiple-line selection biases (weakens) the metallicity dependence
  on SSFR below the SF sequence ($\Delta$ log SSFR$<0$) by
  preferentially eliminating galaxies with lower SSFRs and higher
  metallicities. Dashed lines show median metallicities prior to
  selection by T04 availability and demonstrate the effect of the S/N
  selection. \label{fig:dssfr_z_m10_t04}}
\end{figure*}

\begin{figure*}
\epsscale{1.1} \plotone{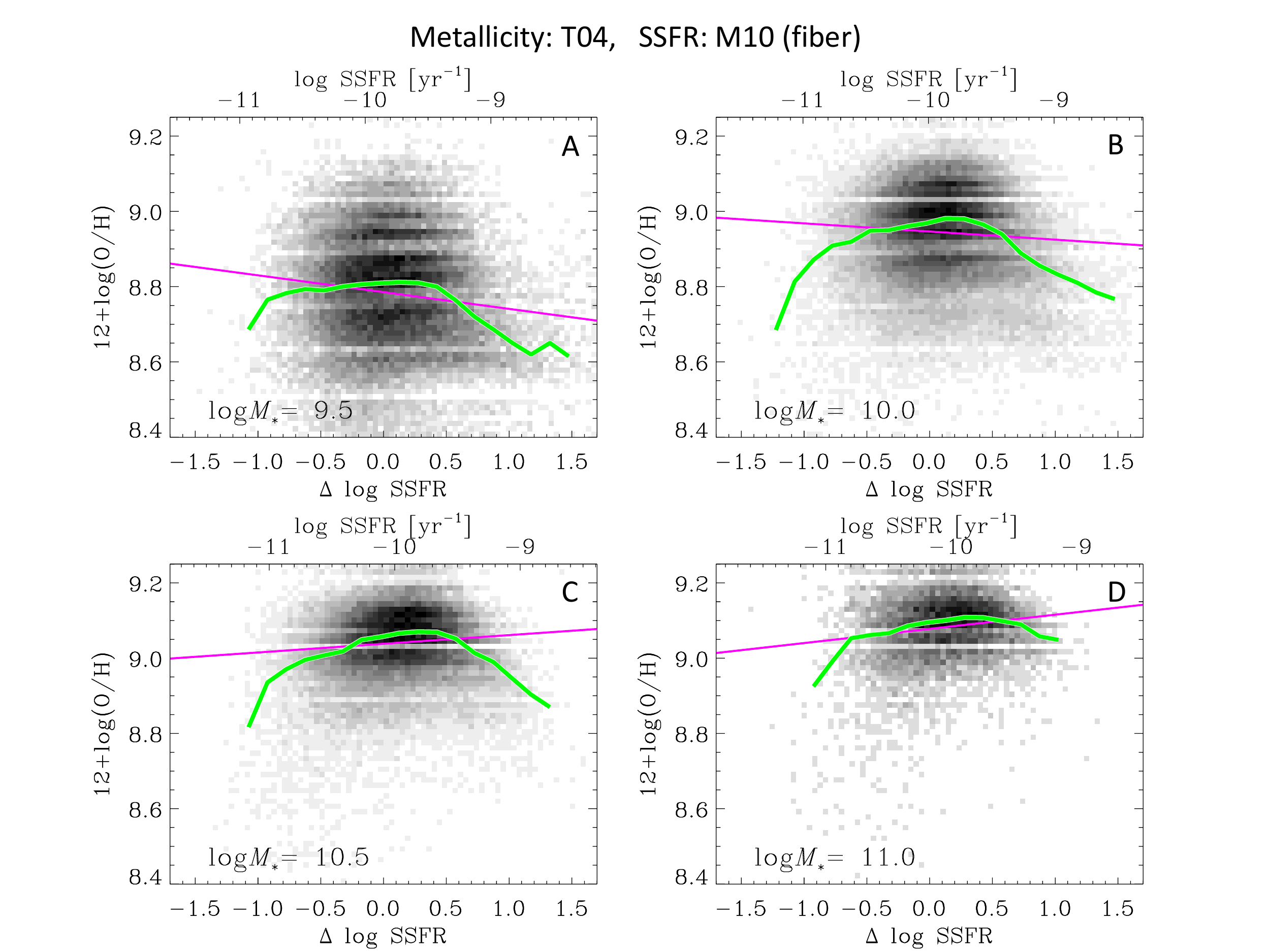}
\caption{Dependence of metallicity on the offset from the star-forming
  sequence in different mass bins, where the metallicity is now from T04,
  while the SSFRs are still derived as in M10. A criterion required
  for M10 metallicities (that the estimates from N2 and R23 methods
  agree) is now dropped. Metallicity trends are weaker compared to
  those with M10 metallicities and are in some cases
  non-monotonic. This is a consequence of both the fact that T04
  metallicities are only available for galaxies that have strong
  S/N ratios in multiple lines {\it and} that the T04 values
  yield weaker trends compared to those computed by M10.
  \label{fig:dssfr_z_t04}}
\end{figure*}

\subsection{Is \mzs\ relation sensitive to the choice of metallicity
  measurements?}

In this section we explore how \mzs\ relation is affected by the use
of alternative metallicity indicators, primarily those based on T04
method. Since T04 metallicities that are available in MPA/JHU catalog
also involve additional selection criteria, we first need to explore
if such criteria {\it alone} affect the dependence on SFR. In the
analysis we will continue to use $Z$--$\Delta$ SSFR plots split by
mass bins.

Given all the practical and theoretical difficulties concerning the
measurement of metallicities (e.g., \citealt{andrews13} and references
therein), it is of critical importance to understand if different
methods of deriving them affect the conclusions regarding the
existence and the character of \mzs\ relation. In all of our analysis up to this
point we have used metallicities derived according to the method of
M10, i.e, an average of estimates based on R23 and N2 calibrations of
\citet{maiolino08}. T04 metallicities, being available as part of the
MPA/JHU catalog, are very often used in the analysis of SDSS samples
and have been used for the characterization of \mzs\ relation
\citep{lara-lopez10,yates12}. However, T04 metallicities are not
available for the entire sample that we considered so far (that was
selected by requiring S/N$(\ha)>25$). Instead, T04 metallicities are
available for galaxies satisfying the condition of having relatively
high S/N ratios in all four BPT lines (S/N ratio $>$7.3, 5.5, 4.5 and
6.0 in $\ha$, $\hb$, [OIII]5007 and [NII]6584, respectively), as
explained in Section 2.2, which is fulfilled for 2/3 of our sample. To
see whether multiple S/N ratio cuts alone bias the trends, we repeat
Figure \ref{fig:dssfr_z_m10}, i.e., we still use {\it M10}
metallicities, but now restricted to those galaxies for which T04
metallicities exist. The resulting trends, in same mass bins as
before, are shown in Figure \ref{fig:dssfr_z_m10_t04}. Comparison with
Figure \ref{fig:dssfr_z_m10} reveals that while the overall trends are
similar, there are important differences. Namely, the median
metallicities {\it below} the SF sequence ($\Delta$ log SSFR$<0$) are
now lower, by up to 0.05 dex. There is no change above the SF
sequence. In other words, the selection based on four emission lines
leads to the preferential removal of galaxies with lower SSFRs and
higher metallicities. The overall result is that the bulk trends
(purple lines in Figure \ref{fig:dssfr_z_m10_t04}) become slightly
weaker.

Keeping in mind the biases introduced by S/N ratio cuts present in T04
sample, we now look at the SSFR trends using the actual T04
metallicities (Figure \ref{fig:dssfr_z_t04}). The results are
remarkably different compared to equivalent plots made with M10
metallicities (Figure \ref{fig:dssfr_z_m10_t04}). In the lowest mass
bin (panel A), where the dependence on SFR was the strongest, it is now
much weaker, with the overall slope of only $\kappa=-0.05$. More
importantly, the sense of the $Z$($M_*$,SFR) (that the more active
galaxies at a given mass should have {\it smaller} metallicities) is
only observed above the star-forming sequence ($\Delta$ log
SSFR$>0.5$, and therefore involves a smaller fraction of
galaxies. Bulk of the galaxies (those within the core of the
star-forming sequence and below it) show no trend at all. Similar
situation persists in higher mass bins. Anti-correlation between
metallicity and SFR is observed only at $\gtrsim0.4$ dex above the SF
sequence, but for the bulk of the galaxies it is not present, and even
turns into a weak {\it positive correlation}. The end result of using
T04 metallicities is that the trends are no longer monotonic, which is
inconsistent with the possibility that the \mzs\ relation is redshift
invariant.\footnote{As explained in Section 2.3, for T04 metallicities
  we use medians of probability distribution functions. However, the
  results stay the same if averages or modes are used instead.}

To disentangle the effects of T04 sample selection that requires high
S/N ratios in multiple lines, from those that are intrinsic to the T04
metallicity method, we have also calculated the metallicities, using
the exact methodology of T04, for galaxies that do not pass T04
cuts. Thus we arrive at the sample that is selected with only the S/N
ratio cut on $\ha$ line. Investigating the metallicity trends with
such sample we still find (plots not shown) that T04 metallicities
lead to much weaker trends against SSFR than M10 metallicities, but
not so much to produce a drop at low SSFRs that would lead to
non-monotonic behavior. This, apparently, is due to the additional
bias of having S/N ratio selections in multiple lines.

Differences between M10 and T04 metallicities in the context of \mzs\ relation
were previously noted by \citet{yates12}, who ultimately preferred the
T04 metallicities. Considering that most studies determine
metallicities using simple methods similar to those used in M10, and
not using the complicated Bayesian fitting of full emission line
spectrum as in T04, it is important to understand if the simple method
of M10 is at fault.

\begin{figure*}
\epsscale{1.1} \plotone{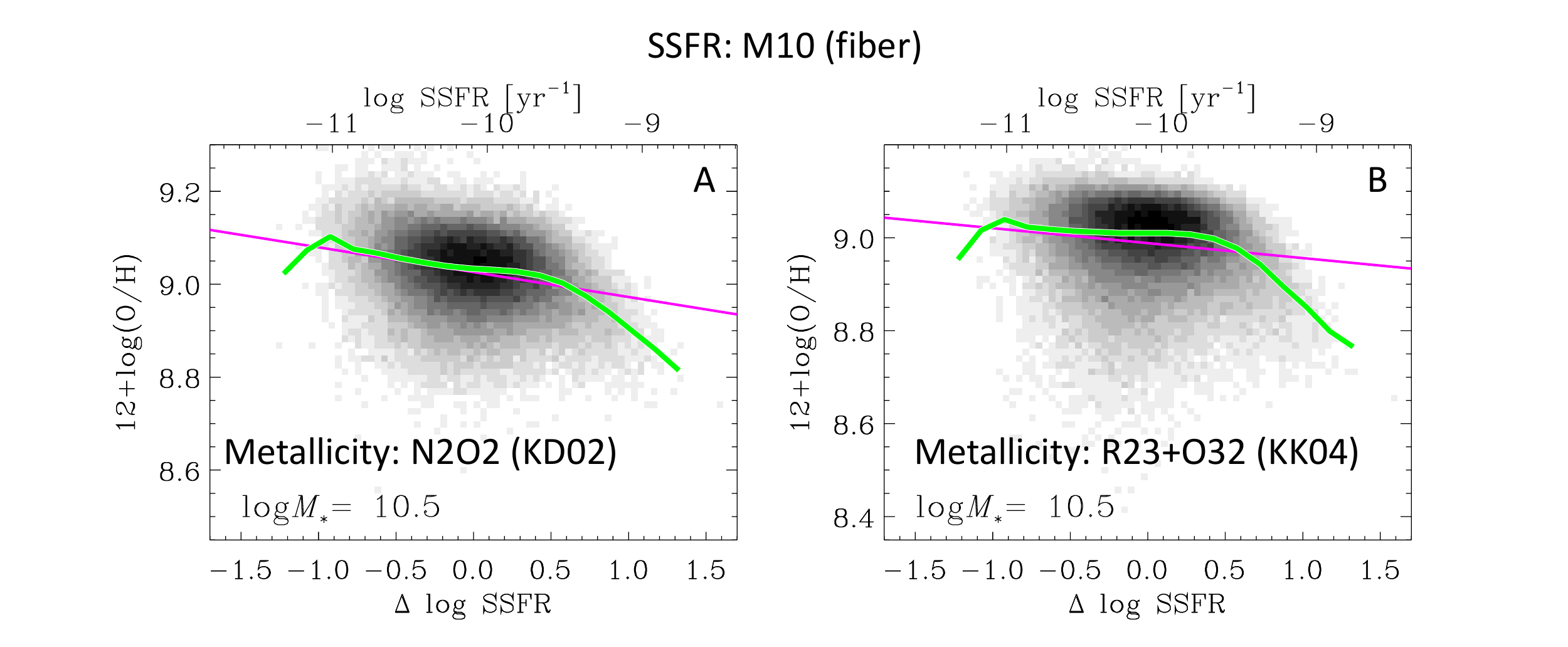}
\caption{Dependence of metallicity on the offset from the star-forming
  sequence at $\log M_*=10.5$, for two alternative metallicity
  methods. Panel A shows N2O2 metallicities calibrated according to
  \citet{kewley02}, and panel B shows metallicities obtained with the
  \citet{kk04} ``best'' method of combining R23 with O32. Both methods
  should be less affected by the changes in the ionization parameter,
  and yet they show trends more similar to those based on M10
  metallicities than based on T04
  metallicities. \label{fig:dssfr_z_other}}
\end{figure*}

One concern regarding the M10 method is that it is based on
semi-empirical calibrations \citep{maiolino08}. \citet{maiolino08}
calibrate various individual line ratios of a sample of SDSS galaxies
against the metallicities obtained from the theoretical calibrations
of \citet{kewley02}. Thus, such calibration will follow the
theoretical models on {\it average}, but not necessarily for parts of
the sample that have properties different from the
average. Specifically, their calibrations may not be valid for
galaxies that have {\it ionization parameters} very different from
what is typical at a given line ratio value. To test the possibility
that the \mzs\ relation based on M10 metallicities is affected by
biases due to the variations of ionization parameter, we also
calculate metallicities using the N2O2 method, which is the least
sensitive to ionization parameter of all simple methods (methods that
employ one ratio of lines) \citep{kewley02}. We show the N2O2
metallicity vs.\ SF trends in Figure \ref{fig:dssfr_z_other}A, but now
only for one mass bin ($\log M_*=10.5$). The general sense of the
trend is the same as it was with M10 metallicities: there is an
anti-correlation for galaxies within the SF sequence, which becomes
stronger above it. The overall slope is even slightly steeper than it
was with M10 metallicities. There is no sign of flat or positively
correlated trends as with T04 metallicities.

In their choice to adopt T04 metallicities as more fiducial,
\citet{yates12} put forward an argument that any method for deriving
metallicities involving nitrogen (such as M10's, which is an average of
N2 and R23) is possibly affected by saturation. To address this
concern, we now consider an oxygen-based (R23) calibration of
\citet{kk04} which at the same time accounts for differences in the
ionization parameter (through its dependence on O32). The results
(again for $\log M_*=10.5$ bin) are shown in Figure
\ref{fig:dssfr_z_other}B. While the median trend within the SF
sequence is now somewhat weaker than for either the M10 or N2O2
metallicities, they do not go away as in the case of T04
metallicities. In other mass bins (plots not shown) both the N2O2 and
the R23+O32 trends are again closer to trends using M10 metallicities
than those using T04 metallicities.

Based on the analysis presented in this section we conclude that the
use of T04 metallicities, even when accounting for the biases due to
having multiple-line selections, produces SFR dependencies that are
much smaller than those produced using conventional line ratio
methods. It should be kept in mind that even among the methods that
yield similar results, the strength of the dependence on (S)SFR will
differ depending on the metallicity method and calibration
\citep{andrews13}. This reiterates the point that the comparisons of
different samples must be based on same metallicity method and
calibration (in addition to using comparable (S)SFR estimates, and
allowing for non-linear trends, as emphasized in previous sections).

Furthermore, in this section we find, as discussed in \citet{mithi},
that the dependence on SFR will depend on the S/N ratio cuts applied to
the lines. This is especially true for galaxies with lower (S)SFRs.
This can be avoided by using a S/N ratio cut only on one hydrogen line
(which can be made sufficiently high so that other lines are also well
measured).

\section{Application of the framework to resolve open questions}

In this section we apply the analysis framework presented in Section
3 to revisit some of the results regarding metallicity's secondary
dependencies presented in recent literature. Specifically, we revisit
the existence of another secondary parameter, galaxy size
\citep{ellison08}, and also explore claims that the character of \mzs\ relation
is different for galaxies of different mass
\citep{yates12,lara-lopez13}, or the claims that the secondary
dependence may be altogether spurious \citep{sanchez13}.

\begin{figure*}
\epsscale{1.1} \plotone{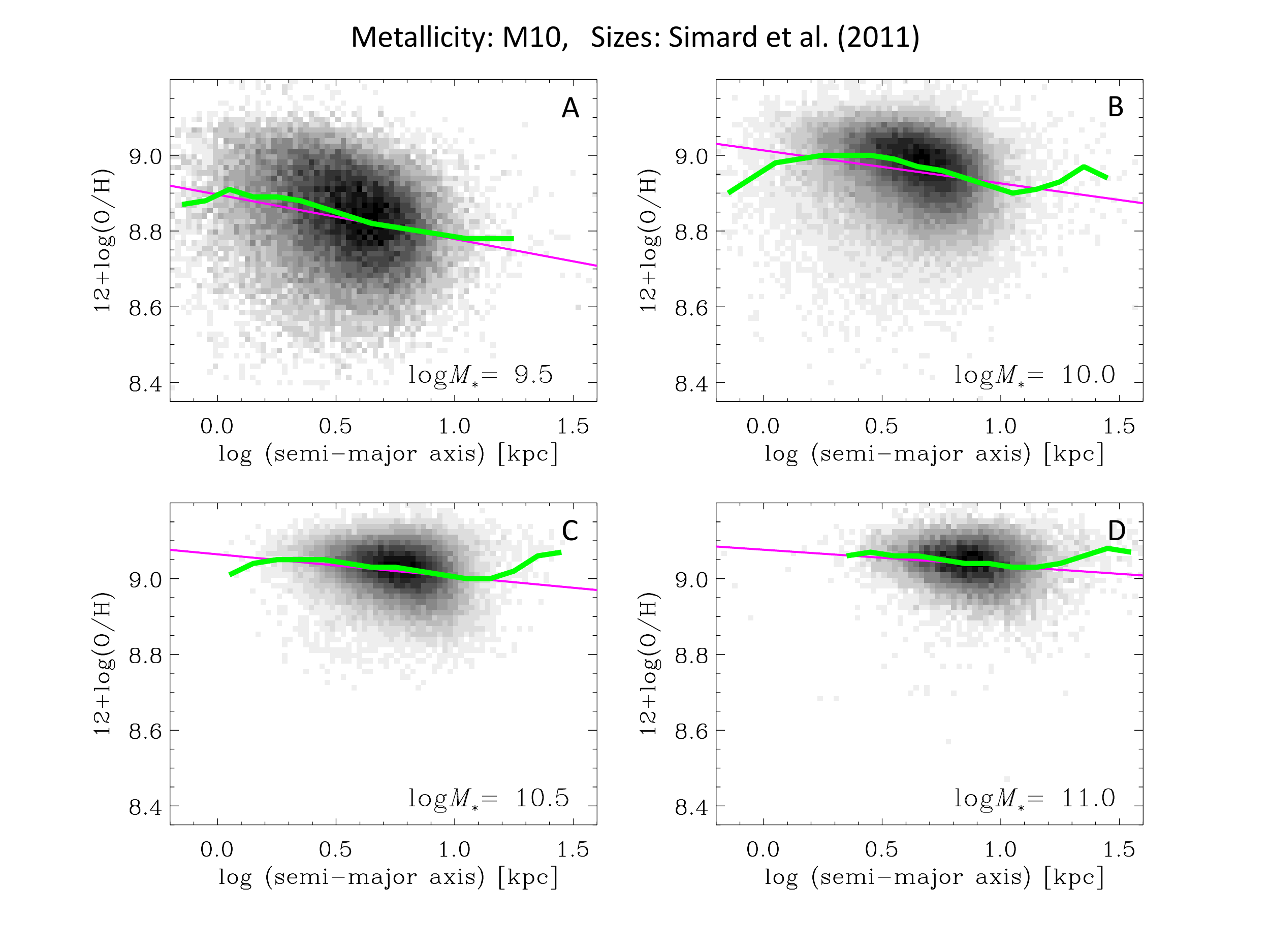}
\caption{Dependence of metallicity on the galaxy size (half-light
  semi-major axis in $g$ band, in kpc) in different mass bins.
  Anti-correlations are present, but the overall trend in median
  metallicities is not as large as those with SSFR (Figure
  \ref{fig:dssfr_z_m10}). \label{fig:rh_z}}
\end{figure*}

\subsection{Is MZR dependent on galaxy size?}

The work that originally drew attention to the dependencies of MZR on
other parameters, \citet{ellison08}, claimed to have found not only
MZR offsets due to SSFR, but also even {\it stronger} offsets
resulting from galaxy physical sizes, in the sense that at a given mass larger
galaxies have on average smaller metallicities. However, subsequent
studies focused exclusively on the SFR aspect, neglecting the question
of galaxy sizes, perhaps because the latter phenomenon had less clear
intuitive interpretation, and because any result involving galaxy
sizes in a sample where the physical scale and the fiber covering
fraction span such wide ranges appears suspect. Furthermore, as
pointed out by \citet{ellison08}, simultaneous presence of both
dependencies leads to the predictions that high-redshift MZRs should
have much smaller evolution than what is observed, so the existence of
a strong size dependence would potentially conflict with the basic
idea of FMR---that it can account for MZR evolution.

Here we revisit the question of MZR's dependence on galaxy size by
applying the methodology introduced in Section 3 (metallicity trends
in 0.5 dex wide mass bins), but substituting SSFR with galaxy size. We
use the same data for galaxy sizes \citep{simard11} as used by
\citet{ellison08}, except that we take semi-major axis half-light
radii in $g$ band (as opposed to $r$), based on single Sersic
fits. The results for our augmented sample, and using M10
metallicities, are shown in Figure \ref{fig:rh_z}. Significant trends
are seen in all mass bins, but they are stronger at lower masses. Bulk
trends (purple fits) are not as strong as those vs.\ SSFR, although
they become more comparable at higher masses. The main difference is
that the dependence vs.\ SSFR has a two-mode behavior such that the
trends above the SF sequence are much stronger, which results in the
full range of median metallicities that is many times greater than the
range of median metallicities due to the size dependence. For example,
at $\log M_*=9.5$ the total span of metallicities due to the size is
$\sim 0.1$ dex, while it is $\sim 0.5$ dex with respect to the SSFR
(Figure \ref{fig:dssfr_z_m10}A). Interestingly, the trend vs.\ size
reverses for very large galaxies (half-light size $>10$ kpc), but
there are very few such galaxies compared to those of smaller size.

One might have a concern that the size dependence is a just
consequence of there being the primary dependence with SSFR. This
would be the case if the size and SSFR were themselves strongly
correlated in the sense that larger galaxies have higher
SSFRs. However, this is not the case. We find (plots not shown) that
the galaxies with higher SSFR on average have smaller sizes than
galaxies with lower SSFRs (up to a factor of 2). Furthermore, the
dependence of metallicity on SSFR persists with the same intensity
even when the galaxies are selected to lie in a very small range of
physical sizes (plots not shown). Therefore, the two phenomena are
independent. Finally, we confirm that the trends of metallicity with
size remain when the sample is restricted to narrow redshift ranges
(0.01), which demonstrates that the observed dependence is not an
artifact of redshift-dependent covering fraction of the fibers.

The conclusion of this section is that the metallicity shows
dependence on galaxy size that is independent from the dependence on
SSFR and is not an artifact of aperture effects, nor of metallicity
gradients within the galaxy (for latter, see
\citealt{ellison08}). However, the total trends are weaker compared to
those with respect to SSFR, so one does not expect that the size
dependence alone will produce much evolution in the MZR and thus
affect the potential invariance of \mzs\ relation . Nevertheless, the
size does appear to be a genuine contributing source of the dispersion
(albeit small) in the MZR and is therefore in need of being tested
with theoretical models.

\subsection{Is there a reversal in Z--SFR anti-correlation at higher
  masses?}

The character of the \mzs\ relation, that at a given mass there is an
anti-correlation between SFR and metallcity, has been brought into
question by \citet{yates12} (and to some extent
\citealt{lara-lopez13}), who find that this dependence {\it reverses}
above $\log M_* \approx 10.2$, such that the higher (S)SFRs are
associated with, on average, {\it higher} metallicities. Such result,
if correct, would basically preclude the ``fundamental'' aspect of the
FMR. Namely, as we go to higher redshifts, and SFRs at a given mass
rise, one would expects, based on \citet{yates12} ``reversal'', that
the metallicities of low mass galaxies would be offset lower compared
to local galaxies (as observed), and that the metallicities of higher
mass galaxies should on average be located {\it above} the local
MZR. No such evolution of MZR has been reported in the current
literature. So either the local relationship between $Z$, $M_*$ and
SFR does not at all hold at other redshifts (is not fundamental), or
there is a problem with the finding that there exists a reversal of
trends.

\begin{figure*}
\epsscale{1.1} \plotone{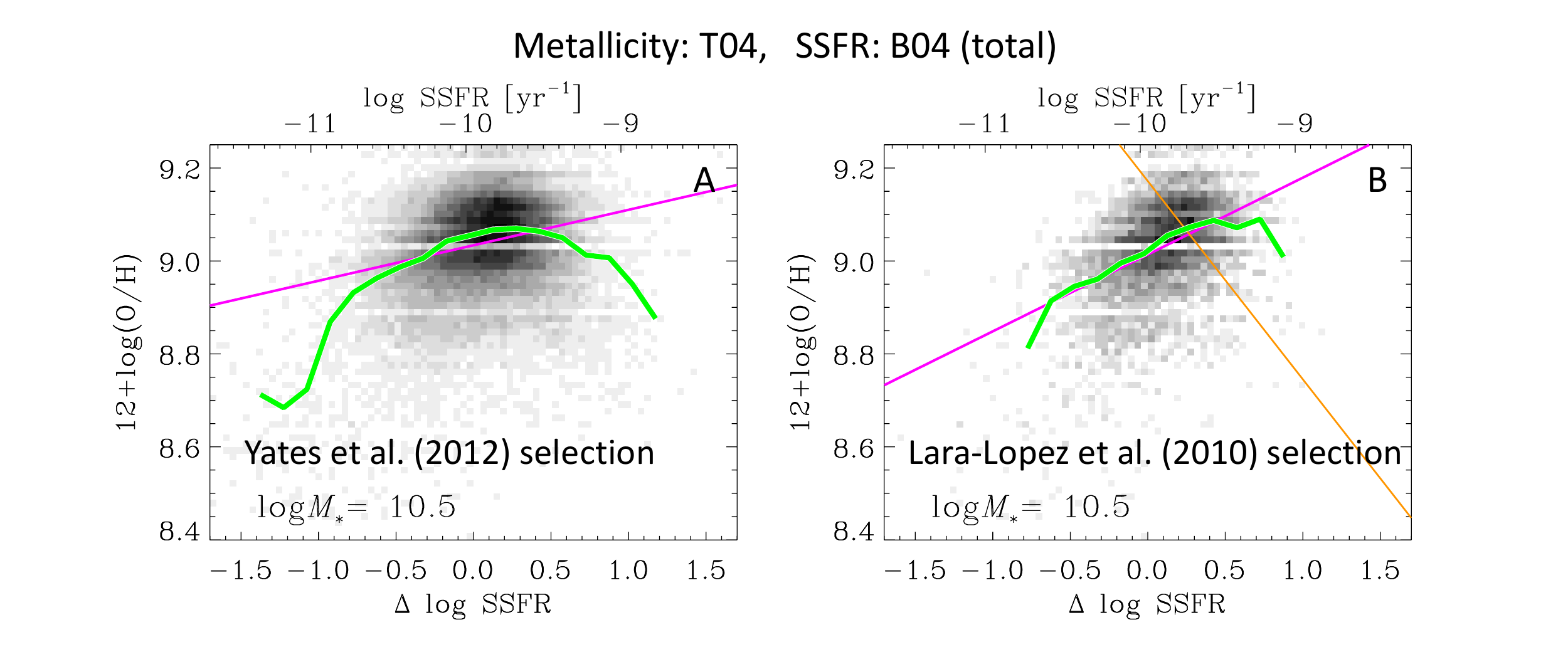}
\caption{Dependence of metallicity on the offset from the star-forming
  sequence at $\log M_*=10.5$, with samples selected as in
  \citet{yates12} (panel A) and in \citet{lara-lopez10} (panel
  B). Both studies utilize metallicities from T04 and total SFRs from
  B04. In panel A the overall trend shows a positive correlation
  between relative SSFR and metallicity, similar to that seen in our
  analysis of T04 metallicities (Figure \ref{fig:dssfr_z_t04}C).
  However, this is the opposite behavior of the anti-correlation seen
  based upon other metallicity measures (Figure
  \ref{fig:dssfr_z_m10}C, \ref{fig:dssfr_z_other}). Even stronger
  positive correlation in panel B can be traced to a lower redshift
  ceiling and higher S/N ratio cuts applied by
  \citet{lara-lopez10}. However, this positive correlation is not
  captured by the \citet{lara-lopez10} ``fundamental plane''
  parameterization of the relation (orange line), which, like the FMR
  of M10, implies that the galaxies follow an
  anti-correlation. \label{fig:dssfr_z_lit}}
\end{figure*}

\citet{yates12} found the reversal using T04 metallicities. They have
also considered, but eventually decided not to trust, the
metallicities determined according to the M10 method, which, even in
their analysis, did not show any evidence of the reversal (their
Figure 1). Our analysis (Section 3.5) has shown that T04 metallicities
intrinsically shows much weaker trends than other metallicity
indicators, which are additionally exacerbated by T04 metallicities
being available only for galaxies that fulfill S/N ratio criteria on
multiple lines, leading to trends that are strongly non-monotonic,
especially at higher masses (Figure \ref{fig:dssfr_z_t04}C, D). Can
this explain the results of \citet{yates12}? In Figure
\ref{fig:dssfr_z_lit}A we show metallicity vs.\ SSFR for mass bin
centered at $\log M_*=10.5$, where, according to \citet{yates12}, the
reversal should already take place. The sample selection used in this
figure follows that of \citet{yates12}, applying S/N>5 cuts in $\ha$,
$\hb$ and [NII]6584, redshift range of 0.005 to 0.25 and the
requirement that fiber captures at least 10\% of $r$-band flux.
Yates et al.\ use aperture-corrected total SFR from B04\footnote{B04
  total SFRs combine the SFR determined in the fiber (Section 2.3)
  with the SFR estimated for the region outside of the fiber. These
  outer-ring SFRs were based on outer broad-band optical fluxes,
  calibrated to match the SFRs of fibers having the same color. As
  such, the total B04 SFRs are emission-line/broad-band hybrids.}.
Indeed, the general sense of the trend in Figure
\ref{fig:dssfr_z_lit}A (purple line) is one of a correlation (and
therefore the reversal compared to the anti-correlation at lower
masses). This is similar to what we have already seen when we
discussed T04 metallicities (Figure \ref{fig:dssfr_z_t04}), but with
even fewer galaxies participating in the anti-correlation part of the
trend, perhaps due to a different type of SFR or somewhat different
sample cuts.

One could argue for reversal being real by noting that a similar
effect may exist in trends of dust extinction vs.\ SFR. Namely,
\citet{zahid13} find that dust extinction $A_V$ (estimated from the
Balmer decrement) in SDSS shows an anti-correlation with SFR at lower
masses, which turns into a positive correlation above $\log M_*>10.2$.
These trends are relatively weak and have a substantial scatter ($\sim
0.5$ mag). For SFR \citet{zahid13} use B04 {\it total} SFRs.
Interestingly, we confirm \citet{zahid13} results, but only when using
total SSFRs. When we instead look at $A_V$ vs.\ {\it fiber} SSFR
(either derived as in M10 or from B04), the mean trends have the same
character irrespective of the mass: dust extinction rises with SSFR,
reaches a peak, and then turns down above the SF sequence. The rising
part of the trend becomes steeper with mass. Given that the Balmer
decrement is determined in the fiber, it is more appropriate to
compare it to the SSFR also measured in the fiber. Thus the results
of \citet{zahid13} may not hold when the dust and the SSFR are measured
in matching regions, possibly because the dust extinction and the SSFR
do not scale alike.

\begin{figure*}
\epsscale{1.1} \plotone{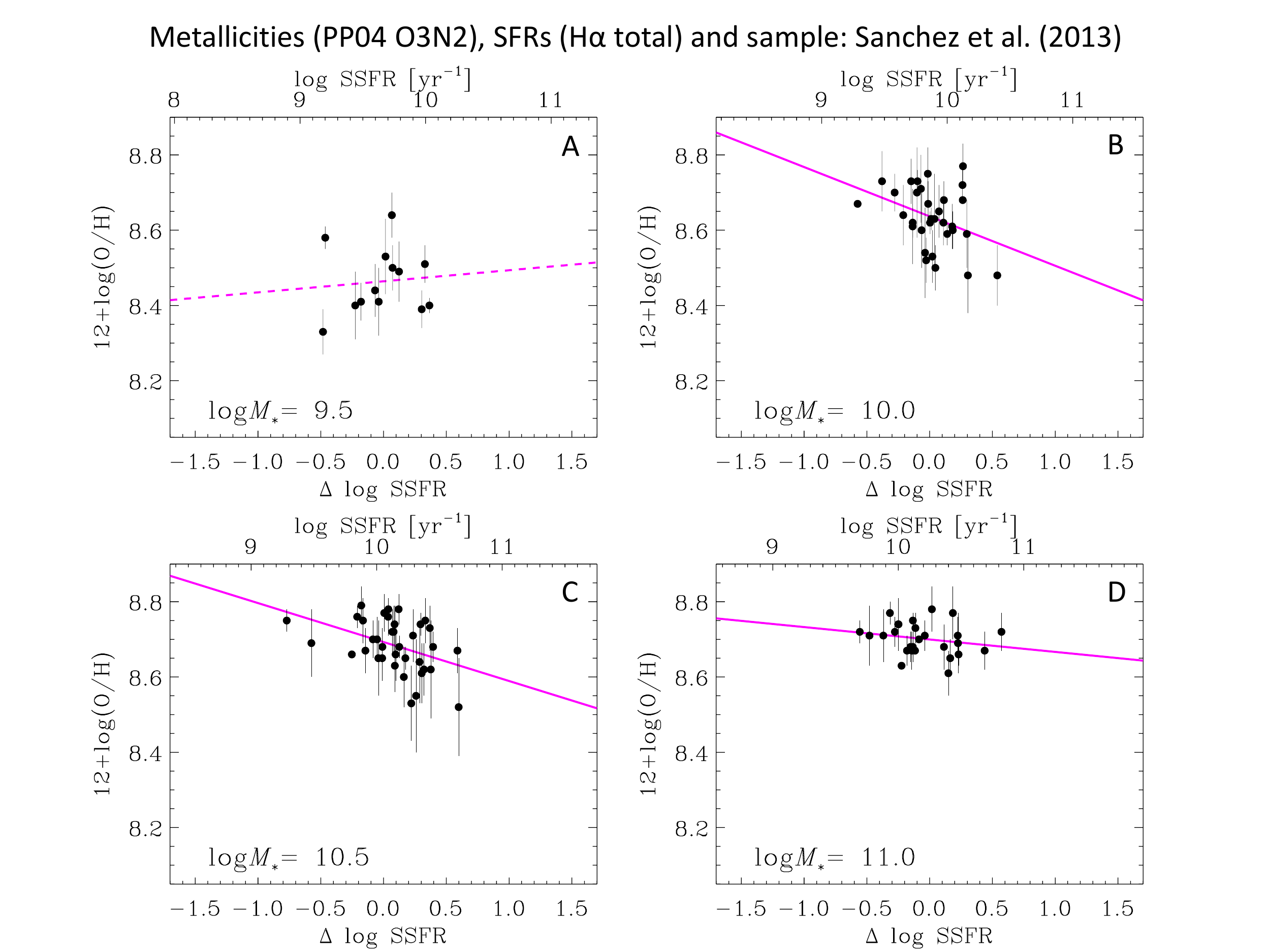}
\caption{Dependence of metallicity on the offset from the star-forming
  sequence for the CALIFA sample of
  \citet{sanchez13}. \citet{sanchez13} observed 150 galaxies with
  integral field spectroscopy and were therefore able to derive
  metallicities and SFRs that better reflect entire galaxies than SDSS
  fiber measures, and should not be susceptible to potential aperture
  biases. Interestingly, we find that (except in the lowest mass bin,
  panel A) statistically significant anti-correlations (purple lines
  show unweighted linear fits) are present, with strengths that are
  comparable to those in SDSS sample. \label{fig:dssfr_z_sanchez}}
\end{figure*}

T04 metallicities were also used in LL10, the other of the two papers
that first reported the relationship between mass, SFR and
metallicity. Like \citet{yates12}, LL10 used total SFRs from B04, but
also a much more restricted redshift range ($0.04<z<0.10$), and very
high ($>8$) S/N ratio cuts in eight emission lines (four BPT lines,
plus [OII]3726, 3729 and [SII]6717, 6731). We replicate LL10 selection
and show the resulting metallicity vs.\ SSFR plot at $\log M_*=10.5$
in Figure \ref{fig:dssfr_z_lit}B. The apparent correlation
(``reversal'') is now even stronger than in \citet{yates12}. This is
the result of lower redshift range which preferentially eliminates
less frequent high-SFR galaxies that would have added weight to the
anti-correlation, and due to the high S/N ratio cuts that tend to
preferentially remove low-SFR galaxies with high metallicities. LL10
do not discuss the reversal in their original paper, but do confirm it
subsequently \citep{lara-lopez13}. However, it must be noted that the
high-mass reversal is entirely at odds with the concept of the
``fundamental plane'' introduced by LL10. The sense of the fundamental
plane is always one of anti-correlation. This can be seen in Figure
\ref{fig:dssfr_z_lit}B, where we plot the locus of LL10 fundamental
plane as the orange line (it is also evident in Figure 13 (top left
panel) in \citealt{lara-lopez13}). Furthermore, the fundamental plane
requires the slope $\kappa$ to be mass-independent (e.g.\ Figure 13
(top left panel) in \citealt{lara-lopez13}), which is obviously not
the case no matter which metallicity or SFR indicator is used.  The
fact that LL10 were able to derive the fundamental plane despite using
T04 metallicities that cause the apparent reversal, is because the
plane was constrained by more numerous lower mass galaxies that
dominate in their sample and for which the general trend is that of
anti-correlation, even using T04 metallicities (Figure
\ref{fig:dssfr_z_t04}A).

The conclusion is that the apparent reversal in metallicity vs.\
(S)SFR, as seen in \citet{yates12}, is primarily the result of using
the available T04 metallicities, and would not be seen using other
metallicity estimates (or even with T04 metallicities if they were
available for a sample not biased by multiple-line S/N ratio cuts).
Furthermore, the reversal would have conflicted with the fundamental
aspect of the local $Z$--$M_*$--SFR relation, because it would no
longer be able to explain the MZR evolution.

\subsection{Is \mzs\ relation merely an artifact of aperture effects?}

One potentially serious limitation of all studies of the relationship
between metallicity, mass and SFR that are based on SDSS data is that
the metallicity measurements come from fiber spectroscopy, which
covers part of the galaxy in a way that is redshift and galaxy-size
dependent. \citet{sanchez13} made efforts to address this concern by
observing a sample of 150 local ($z<0.03$) galaxies using an integral
field spectrograph PMAS/PPAK mounted on Calar Alto 3.5 m, as part of
the CALIFA survey. In addition to the measurements of resolved HII
regions, \citet{sanchez13} determine global estimates for galaxy
metallicity in a physically motivated way (at one effective radius),
as well as the total SFRs based on $\ha$. Such sample, even though
relatively small, but being free from aperture effects, could prove
essential in either strengthening or weakening the status of the
\mzs\ relation. The analysis performed on global measures by \citet{sanchez13}
concluded that no dependence of MZR on SFR existed. Furthermore, they
tentatively explained the apparent presence of this dependence in SDSS
data to be due to the aperture affects (their Appendix).

Here we reanalyze \citet{sanchez13} data using our preferred
methodology: metallicity vs.\ relative SSFR in individual mass bins.
The results are shown in Figure \ref{fig:dssfr_z_sanchez}). Except in
the lowest mass bin (panel A), the anti-correlation between
metallicity (determined by \citealt{sanchez13} using \citet{pp04}
calibration of O3N2 method) and SSFR (i.e, the offset from the SF
sequence as derived with \citealt{sanchez13} data) is convincingly
present (purple lines show linear unweighted fits). In mass bins
centered at $\log M_*=10.0$ and $10.5$ there is only 5\% and 3\%
probability that the anti-correlation is due to chance (obtained using
bootstrap resampling; similar results, 3\% and 0.1\%, are obtained
when measurements are perturbed within the error bars). The positive
correlation in the lowest mass bin is not statistically significant
(there is a 37\% probability that it is due to chance), but it is
incompatible with very strong anti-correlation expected at those
masses. We point out that \citet{sanchez13} sample is incomplete at
those masses, and the apparent lack of anti-correlation could
potentially be due to the apparent size selection present in CALIFA
dataset \citep{walcher14}. We believe that the reason why this
dependence was not detected in the analysis of \citet{sanchez13} was
because the sample was not split by stellar mass (their Figure 4,
bottom right panel). Furthermore, in a small sample that lacks extreme
star-formers, the trend in metallicities will be relatively modest and
therefore difficult to spot on a mass-metallicity plot color-coded by
SFR (their Figure 4, lower left panel).

We conclude that it is very encouraging that the measurements that
avoid the issues of SDSS fibers confirm the MZR dependence on SSFR,
and in a degree that is comparable to that using more extensive SDSS
data.

\section{Discussion: Implications for theoretical studies}

Chemical enrichment of galaxies and its change through
cosmic time is the result of an intricate interplay between
star formation (turning gas into stars), stellar evolution (releasing
enriched gas into the ISM), regulation of SF (different forms of
feedback), galaxy-scale outflows (possibly related to feedback
processes), as well as gas accretion (from intergalactic medium (IGM),
recycled outflow gas, or from gas-rich mergers). Each of these
individual processes are themselves not fully understood.

There are two main aspects of chemical enrichment: that of the stars,
and of the ISM (typically of gas in photo-ionized HII regions)). Both
are usually expressed in terms of a metallicity. Stellar metallicities
are more representative of the sum record of the history of metal
enrichment, while gas metallicities are more reflective of the current
level of chemical enrichment. The study of stellar metallicities of
individual galaxies requires good absorption-line spectroscopy, hence
it is mostly limited to low redshifts
\citep{gallazzi05}, with pioneering efforts at intermediate and high
redshifts currently under way \citep{sommariva12,gallazzi14}.

Theoretical efforts first focused on trying to reproduce the local MZR
and its evolution with redshift, and were only more recently modified
to address the dependence on SFR. In principle, the existence of MZR can
be explained in the context of a closed-box model, simply as the
consequence of ``downsizing'', a scenario in which more massive
galaxies have produced most of their stars and metals early in the
history of the universe (e.g.,
\citealt{garnett02,savaglio05}). However, simple closed-box scenarios
violate numerous constraints, including Milky Way G- and M-dwarf
metallicities~\citep[e.g.,][]{woolfwest12}. Thus, building on early
ideas by \citet{dekelsilk86}, T04 proposed a model in which galactic
winds (outflows), which are responsible for the removal of metal-rich
gas from SF regions, are more efficient in low-mass galaxies, leading
to the observed MZR. More recent analytic models typically require
both inflow and outflow to match galaxy
metallicities~\citep[e.g.,][]{dalcanton07,peeples11}.

Hydrodynamic galaxy formation simulations that include strong feedback
have been able to reproduce the MZR.  \citet{brooks07} used
high-resolution zoom disk galaxy simulations to argue that the MZR
faint-end slope is primarily set by the lowered efficiency of
converting gas into stars in the ISM (due to supernova feedback), as
opposed to ejecting metals.  \citet{finlator08} and \citet{dave11}
used lower-resolution cosmological simulations to argue that outflows
lowered the efficiency of converting infalling (not ISM) gas into
stars, and \citet{dave12} showed that this can be effectively
parameterized in a simple analytic framework that predicts $Z\propto
\eta^{-1}$ in small galaxies, where $\eta$ is the mass loading factor.
The observed faint-end MZR of $Z\propto M_*^{0.3-0.4}$ then implies
$\eta\propto M_*^{-1/3}$, consistent with momentum-driven
winds~\citep{murray05,oppenheimer08}.  More recent cosmological
simulations favor a steeper scaling of $\eta$ to reproduce the stellar
mass function, but this results in an MZR that is too
steep~\citep{dave13}.  Over-enriching outflows, as suggested by
data~\citep[e.g.,][]{heckman00} makes the problem worse, and indeed the
Illustris simulation employs {\it under-}enriched outflows, which is
difficult to justify physically but improves agreement with
data~\citep{vogelsberger14}.  Recent zoom simulations that include
$H_2$-based star formation can reproduce the MZR and stellar mass data
via a combination of outflows driving out gas and metals together with
a reduced ISM star formation efficiency owing to lower
metallicities (L. Christensen, in prep.)  Clearly the physics that sets
the MZR shape is not fully sorted, but successful models commonly
invoke increasingly stronger outflows to low-masses, with saturation
at high masses where outflows become ineffective.  Regarding the MZR
evolution, outflows are necessary to explain slow enrichment by
$z\sim2$ \citep{finlator08}, but the overall increase of metallicity
at a given mass is due to the accreted gas becoming more metal rich
\citep{dave11}.

Cosmological simulations concurrently predict the \mzs\ relation.
\citet{dave11} showed that projecting the simulated galaxies onto the
FMR plane of M10 indeed lowered the scatter, though not by quite as
much as expected from M10.  The trend qualitatively arises in these
models because pristine infall both increases the gas content to
stimulate star formation, while reducing the gas-phase metallicity.
Galaxies thus fluctuate around the ``equilibrium" MZR owing to
fluctuations in the infall rate (such as mergers).  \citet{finlator08}
showed that the MZR scatter is thus set by the timescale to return to
the ``equilibrium" MZR, and in simulations when this dilution
timescale became greater than the halo dynamical time, the MZR scatter
blew up.  Such ideas were encapsulated in the analytic ``gas
regulator" model of \citet{lilly13}, in which the metallicity is
determined instantaneously by gas consumption timescale ($\epsilon$),
mass loading of wind outflow ($\lambda$) and the specific SFR. The
\mzs\ relation emerges in this model if $\epsilon$ and $\lambda$ are
mass-dependent. Furthermore, in gas regulator model the \mzs\ relation
is redshift-invariant if $\epsilon$ and $\lambda$ are themselves
constant with time. The model of \citet{lilly13} was generalized by
\citet{pipino14} to allow for an evolving efficiency of SFR from
inflowing gas, and was thus able to provide a very good match to both
the MZR and the \mzs\ relation.  \citet{zahid14b} forwarded a related
empirical model in which gas infall dilutes the existing metallicity,
and hence argued the MZR evolution is fundamentally governed by a
relation between metallicity and gas-to-stellar mass ratio.  However,
inflow fluctuations are not the only viable explanation.
\citet{dayal13} suggested instead that the \mzs\ relation arises
because higher SFR galaxies have stronger outflows that eject more
metals.  A review of recent theoretical ideas related to gas accretion
and its impact on the MZR can be found in \citet{sanchezalmeida14}.

Results of our study have a number of implications for theoretical
efforts. We have seen that the exact character of \mzs\ relation will
change depending on the metallicity, and to some extent, the SFR
indicator. Therefore, theoretical results should not be expected to
reproduce details of any empirical relation, but instead should lie in
the range of empirical estimates. The strength of the correlations
will be affected by the accuracy of measured quantities, in particular
the ability to accurately identify galaxies with high SFRs. Therefore,
it is recommended that the results of simulation also include
realistic effects of observational errors. In the same way in which
the mass-binned metallicity vs.\ SSFR plots were shown to be a useful
framework to characterize the \mzs\ relation empirically, so it is a
recommended way to show theoretical predictions and compare them to
the observations.

Our results further challenge models to produce not only the
dependence on SFR
, but also make it stronger for lower-mass
galaxies. Also, what we find to be a consistent feature among galaxies
of $\log M_*\lesssim 10.5$ is the change in strength of SFR-dependence
above the SF sequence. Theoretical work has yet to address this. Our
suggestion is that it reflects the change in the mode of
star-formation above the SF sequence, possibly in relation to galaxy
interactions. If mergers are common in this regime, one expects lower
metallicities (and therefore stronger $Z$--SSFR trends) simply due to
the progenitor bias. Note that a late-stage 1:1 major merger with the
final mass $\log M_*=10$ will still have the metallicity of a $\log
M_*=9.7$ galaxy because it hasn't had time to enrich its gas yet.  The
factor of two in mass around $\log M_*=9.7$ corresponds to roughly a
0.1 dex shift in metallicity (e.g., Figure 2), which would explain a
some of the difference between purple and green lines in Figure
\ref{fig:dssfr_z_m10}B.

Finally, we confirm that the dependence of metallicity on galaxy size
is real, and independent of the trends with SFR. One possibility is
that the galaxies that are larger than what is typical for their mass
are undergoing higher rates of accretion onto the disk, which is then
reflected in the overall reduction of the metallicity.

\section {Conclusions}

The aim of this work was to establish a more physically motivated
non-parametric framework for the study of the \mzs\ relation, and to
apply the methodology to understand the origins of conflicting results
regarding the characterization of the local relation.  We demonstrate
that such a non-parametric framework is needed to accurately
determine whether the MZR has secondary dependencies on other
parameters, and whether local and higher-redshift samples can be
described with a single ``fundamental" metallicity relation. Here, we
have sought to provide a more coherent picture of the empirical
properties of the local \mzs\ relation, and summarize our results as
follows:

\begin{enumerate}

\item A more physically motivated second parameter for the $M_*--Z$
  relation is the {\it relative specific} SFR, i.e., the level of star
  formation compared to what is typical for a galaxy of that mass,
  rather than the {\it absolute} (S)SFR.  The relative SSFR represents
  the offset from the star-forming sequence (the ``main sequence'')
  along the SSFR axis. Selection of galaxies with relatively high SSFR
  better mimics high-redshift selection. The use of {\it specific}
  SFR, as opposed to SFR, also has the advantage that the measurement
  of star formation within SDSS spectroscopic fibers is physically
  meaningful, since it represents the intensity of star formation in
  the same region where the metallicity is measured.
   We caution that absolute fiber SFRs, which were used in the
  \citet{mannucci10} formulation of FMR are strongly distance
  dependent and cover on average of only 25\% of the total SFR.

\item Following from conclusion 1, our preferred framework for the
  study of $M_*--Z$ secondary dependencies and for investigating
  whether the same trends apply at different redshifts (i.e., the
  FMR), consists of plotting the metallicity against the relative SSFR
  as an independent parameter, but restricted to galaxies of a
  certain, relatively narrow mass range (0.2--0.5 dex) at a time. This
  method is free from assumptions of the parametrization of \mzs\
  relation, and exposes important details of the relationship which
  are not captured by previous parameterizations (e.g., a plane, or
  projection of minimal scatter; see Conclusions 3, 4, 5;
  \citealt{mithi}).  All figures in this paper, beginning with Figure
  4, apply and illustrate this method.

\item We confirm that metallicity's dependence on SSFR is weaker for
  more massive galaxies, becoming very weak or absent above $\log
  M_*\approx10.5$ \citep{ellison08}.

\item The secondary dependence on SSFR has a markedly different
  character for intense star formers (with SSFR $\gtrsim$0.6 dex above
  the star-forming sequence) than for ``normal'' star-forming galaxies
  (those in the core or below the sequence). The trend indicates a
  possible two-mode behavior, which can be parameterized with a broken
  linear fit of \oh\ vs.\ log SSFR, with slopes, $\kappa_{\rm high}$
  and $\kappa_{\rm low}$.  The galaxies above the SF sequence have a
  much stronger dependence on the SSFR, which notably, have similar
  $\kappa_{\rm high}$ regardless of the mass. However, as the mass
  increases, a smaller and smaller percentage of galaxies belongs to
  this group of high SSFR galaxies.

\item Conclusion 4 implies that characterizing the relation between mass,
  metallicity and SFR with a flat plane
  \citep{lara-lopez10,lara-lopez13}, or with a projection that
  minimizes the scatter (Eq.\ 5 of \citealt{mannucci10}), forces the
  trends to be identical at different masses and/or at different
  SSFRs, which is not the case. The use of such descriptions of the
  local \mzs\ relation, which would be dominated by weaker trends of
  the majority of local galaxies, could lead to incorrect predictions
  for high-redshift samples, off by 0.2 dex or more in metallicity
  \citep{maier14}.

\item Contrary to common wisdom, accounting for SSFR dependence has a
  modest effect on the reduction of scatter in metallicities---it is
  at most 20\% at the lowest masses, and down to 0\% at higher masses,
  confirming the results of \citet{ellison08}. This is in contrast to
  the reduction of the scatter of median-binned values, which is more
  dramatic \citep{mannucci10}. In other words, the \mzs\ relation
  cannot be thought of as a thin surface. Furthermore, the remaining
  scatter is still higher than the formal metallicity measurement errors,
  suggesting that other parameters may be more closely related to
  metallicity than the (S)SFR.

\item For the majority of galaxies that do not have very high SSFRs,
  the strength of metallicity's dependence on SSFR ($\kappa_{\rm
    low}$) is similar when using fiber SSFRs (based on emission lines,
  particularly $\ha$) or the total SSFRs (based on integrated fluxes),
  eliminating concerns that the SFR dependence is due to a spurious
  correlation between metallicity and emission-line based SFR. For
  intense star-formers ($\gtrsim$0.6 dex above the star-forming
  sequence) $\ha$ SSFRs produce stronger trends than mid-IR SSFRs,
  which could due to the shorter timescales that t$\ha$ is sensitive
  to, or the fact that it measures SFR in the same region in which the
  metallicity is measured (fiber).

\item The character and the strength of trends of metallicity vs.\
  SSFR are sensitive to signal-to-noise ratio selection cuts applied
  to the emission lines. The least biased method is to only select on
  the S/N of a single Balmer line. To ensure that usable metallicity estimates
  are obtained, this cut can be relatively high
  \citep{mannucci10}. Applying cuts to multiple lines preferentially
  removes high-metallicity galaxies with lower SSFRs, effectively
  leading to weaker metallicity trends \citep{mithi}.

\item The choice of metallicity indicator affects the strength of the
  $Z$ vs. SSFR trends, but an anti-correlation is always observed:
  higher SFRs at a given mass on average have lower metallicities. The
  exception is for metallicities derived using the method of
  \citet{tremonti04} (T04, which show non-monotonic behavior, with
  average metallicities decreasing both above and below the SF
  sequence, especially at higher masses. The behavior arises both from
  the signal-to-noise ratio cuts applied to multiple lines for which
  T04 metallicities are available, (Conclusion 8), and from the fact
  that T04 values yield weaker metallicity trends. More work is needed
  to establish the root causes of such differences with respect to
  other methods. The ``reversal'' reported in some recent studies,
  i.e., that the trend of metallicity vs.\ (S)SFR becomes {\it
    positively} correlated for high-mass galaxies can be attributed to
  the use of T04 metallicities in these studies. The reversal is not
  consistent with the concept on an FMR as it would predict that the
  high-redshift MZRs are offset below the local MZRs at low mass and
  offset {\it above} the local MZRs at high mass.

\item Application of our methodology shows that the dependence of
  metallicity on SSFR is present in the CALIFA dataset, which is based
  on integral field spectroscopy for local galaxies, whereas
  \citet{sanchez13} reported no significant secondary dependence on
  SFR, and concluded that the \mzs\ relation is an artifact of
  spectroscopy aperture biases.  We show that the CALIFA data have a
  dependence on the SSFR that is broadly consistent with the relation
  followed by galaxies in the SDSS, except at the lowest masses, where
  the available CALIFA data are few and show no clear dependence.
  
\item We confirm that metallicity has a secondary dependence on
  galaxy size (half-light semi-major axis), as originally found by
  \citet{ellison08}, and that it is independent of the
  dependence on (S)SFR, and also not the result of aperture effects. At
  masses above $\log M_*\gtrsim 10$ the strength of this correlation
  is similar to the dependence with respect to SSFR, but the total
  extent of the median metallicities due to the galaxy size is
  smaller than due to SSFR. The end result is that the dependence on
  galaxy size is secondary to that on SSFR and is therefore less
  relevant in the evolutionary context, but still in need of a
  theoretical explanation.
\end{enumerate}

The non-parametric analysis framework presented here will be used to
evaluate whether the relation defined in the local universe by SDSS galaxies
also describes galaxies at higher redshift (i.e., whether it
is ``fundamental'') in future work (Salim et al.\ 2015, in prep.).  

\acknowledgments We gratefully acknowledge efforts to design,
construct and operate SDSS, {\it GALEX} and {\it WISE}, and produce
and disseminate their data products. This work was supported through
NASA ADAP award NNX12AE06G. We thank Steven Janowiecki for his help in
assembling some of the datasets and Molly Peeples, Robert Yates and
Liese van Zee for discussions regarding their work.


\end{document}